\documentclass[%
 reprint,
nofootinbib,
 amsmath,amssymb,
 aps,
]{revtex4-2}

\usepackage{graphicx}
\usepackage{dcolumn}
\usepackage{bm}
\usepackage{slashed,enumitem}
\usepackage{xcolor,wasysym}
\usepackage[colorlinks=true,allcolors=blue]{hyperref}

\definecolor{defgolden}{rgb}{0.95,0.7125,0.}
\definecolor{deflightblue}{rgb}{0.363898,0.618501,0.782349}
\definecolor{defgreen}{rgb}{0,0.6,0}
\definecolor{defpurple}{rgb}{0.5,0,0.5}
\definecolor{defcyan}{rgb}{0.,0.5,0.5}
\definecolor{defbrown}{rgb}{0.6,0.3,0.1}
\definecolor{defdarkgray}{rgb}{0.35,0.35,0.35}
\definecolor{deflightgray}{rgb}{0.65,0.65,0.65}

\begin{document}


\title{
Why detect forward muons at a muon collider
}

\author{Maximilian Ruhdorfer}
 \email{m.ruhdorfer@stanford.edu}
\affiliation{%
 Laboratory for Elementary Particle Physics, Cornell University, Ithaca, NY 14853, USA\\
 Stanford Institute for Theoretical Physics, Stanford University, Stanford, CA 94305, USA
}%


\author{Ennio Salvioni}
 \email{e.salvioni@sussex.ac.uk}
\affiliation{
 Department of Physics and Astronomy, University of Sussex, Sussex House, BN1 9RH Brighton, UK
}%
\author{Andrea Wulzer}
\email{andrea.wulzer@cern.ch}
\affiliation{%
{{Institut de F\'{\i}sica d'Altes Energies (IFAE), The Barcelona Institute of Science and Technology (BIST),
Campus UAB, 08193 Bellaterra, Barcelona, Spain}} and \\
{{ICREA, Instituci\'o Catalana de Recerca i Estudis Avan\c{c}ats, 
Passeig de Llu\'{\i}s Companys 23, 
08010 Barcelona, Spain}}
}%



\begin{abstract}
\noindent{}We survey the opportunities offered by the detection of the forward muons that accompany the creation of neutral effective vector bosons at a muon collider, in different kinematic regimes. Vectors with relatively low energy produce the Higgs boson and the extended muon angular coverage enables studies of the Higgs properties, such as the measurement of the inclusive production cross section and of the branching ratio to invisible final states. New heavy particles could be produced by vectors of higher energy, through Higgs portal interactions. If the new particles are invisible, the detection of the forward muons is essential in order to search for this scenario. The angular correlations of the forward muons are sensitive to the quantum interference between the vector boson helicity amplitudes and can be exploited for the characterisation of vector boson scattering and fusion processes. This is illustrated by analysing the $CP$ properties of the Higgs coupling to the $Z$ boson. \\[3pt]
\noindent{}Our findings provide a physics case and a set of benchmarks for the design of a dedicated forward muon detector.
\end{abstract}

\maketitle


\section{\label{sec:intro}Introduction}

The International Muon Collider Collaboration (IMCC) is pursuing the technical feasibility of a muon collider with centre of mass energy of 10~TeV or more and with high luminosity~\cite{Delahaye:2019omf,Accettura:2023ked}, implementing the recommendation of the 2020 update of the European Strategy for Particle Physics and the European Roadmap for Accelerator R{\&}D~\cite{Adolphsen:2022ibf}. The Swnomass and P5 processes recently recommended~\cite{Narain:2022qud,P5:2023wyd} a direct United States involvement in the study. 

Muon collider physics, experiment and detector are also being investigated extensively (see Refs.~\cite{Accettura:2023ked,AlAli:2021let,Black:2022cth} for an overview) with a twofold aim. On the one hand, the goal is to establish a preliminary design and to study the performances of a detector operating in the novel conditions of muon collisions, as well as to identify the required detector technologies. On the other hand, the already broad physics case of the muon collider needs to be further expanded by exploring novel opportunities. Physics studies with direct impact on the detector performance specification requirements, like those we report here, are particularly desirable.

In this paper we survey the opportunities offered by the detection of forward muons at the 10~TeV muon collider, with the purpose of providing a physics case and a set of benchmarks that define performance targets for the design of the required dedicated detector. 

At a muon collider, the detector needs to be screened from the radiation that originates from the decay of the muons in the collider ring. This requires the installation of conical absorbers in the interaction region. In the current design, this limits the angular acceptance of the general-purpose (main) detector to 10~degrees from the beam line, i.e.~to a pseudo-rapidity $|\eta|<2.44$. Future design work might improve the acceptance, bringing it closer to that of future $e^+e^-$ colliders~\cite{Fuster:2009em,Aplin:2013oca}, with $|\eta|<4$ being the aspirational IMCC target~\cite{InternationalMuonCollider:2024jyv}. However, the main detector will surely not have access to the angular region below few degrees (i.e., $|\eta|\sim 4\,$-$\,5$), which can instead be covered at proton colliders such as the LHC by the forward calorimeters. 

Unlike electrons and protons, high-energy muons are penetrating particles. If emitted with a small angle, they cross the absorbers and other elements of the collider and could be detected by a dedicated system installed outside, or possibly partially inside, the conical absorbers. The theoretical possibility of extending the angular coverage---though only for muons---to a pseudo-rapidity $|\eta|\sim6$ or more has been known for a long time. The forward muon detector is included in the muon collider \texttt{DELPHES} card~\cite{delphes_card_mucol}, even if a quantitative assessment of its expected performance is not yet available. By contrast, significant literature exists~\cite{Ruhdorfer:2019utl,Ruhdorfer:2023uea,Li:2024joa,Forslund:2022xjq,Forslund:2023reu,Bandyopadhyay:2024plc,Barducci:2024kig,Frigerio:2024pvc,Bandyopadhyay:2024gyg} on the benefits of detecting forward muons for Higgs physics and searches for particles Beyond the Standard Model (BSM).

The generic motivation~\cite{Ruhdorfer:2023uea} for forward muon detection is to further improve and expand the strong physics opportunities associated (see for instance Refs.~\cite{Costantini:2020stv,Han:2020pif,Buttazzo:2020uzc,Buttazzo:2018qqp,Liu:2021jyc}) with the study of reactions initiated by effective vector bosons emitted collinearly by the incoming muons. The emission of the neutral $Z$ boson is accompanied by a muon with absolute rapidity in the typical range $3\lesssim|\eta|\lesssim6$, which is beyond the coverage of the main detector. Accessing muons in this rapidity range through a dedicated detector would offer new handles to investigate Vector Boson Scattering (VBS) and Fusion (VBF) processes. In particular, it enables tagging the emission of the neutral vector for signal selection and measuring otherwise inaccessible properties of the scattering process. The studies presented in this paper are selected to illustrate these capabilities.

In the context of Higgs physics, forward muon detection would enable the precise measurement of the Higgs to invisible branching ratio~\cite{Ruhdorfer:2023uea} and of the inclusive production cross section, enabling in turn the absolute determination of the Higgs couplings~\cite{Li:2024joa}. We revisit these studies with slightly improved background simulation and selection cuts, confirming the sensitivity projections of Refs.~\cite{Ruhdorfer:2023uea,Li:2024joa}.

In the context of BSM particle searches, a clear case is when the new particles are invisible and they are produced by the collision of vector bosons, so that the forward muons are the only detectable objects in the final state. The VBF production of invisible BSM particles is a signature of Higgs portal models. These extensions of the SM can explain the observed abundance of dark matter (see e.g.~Ref.~\cite{Argyropoulos:2021sav} for a review) or emerge in solutions to other fundamental problems, such as the baryon asymmetry of the Universe and the naturalness of the weak scale~\cite{Craig:2014lda,Curtin:2014jma}. By the Goldstone boson equivalence theorem, the portal coupling with the Higgs field is equivalent to an interaction of the invisible particles with the longitudinally-polarised (i.e., zero-helicity) massive vector bosons. The latter interaction is responsible for VBF production of the new particles.

Reference~\cite{Ruhdorfer:2019utl} studied the muon collider sensitivity to renormalisable and derivative Higgs portal models, but neither included some of the relevant backgrounds nor a modelling of the beam energy spread and of the finite resolution of the muon detector. We find that these effects reduce the sensitivity significantly, especially for heavier mass of the putative invisible BSM particle. This is due to a less effective background rejection in the kinematic configuration where the muons lose a significant fraction of their energy, as is needed in order to produce the heavy particles.

It is also possible to exploit the forward muons as a probe of the splitting process that creates the effective vector bosons. This gives access to properties of the colliding vectors and enables extracting more information on the vector boson scattering process. A particularly interesting observable that we consider in this paper is the azimuthal angle of the forward muons. Its distribution is sensitive to the quantum interference between vectors of different helicities, while on the contrary such interference effects cancel out in any observable that is inclusive (integrated) over the angles.

The ability to measure the interference of helicity amplitudes opens up a plethora of opportunities for the characterisation of VBS and VBF processes. As a simple but important illustration, we study the determination of the $CP$ structure of the Higgs boson coupling to the $Z$. A $CP$-odd component can be observed in the neutral VBF Higgs production process only by accessing the interference between different vector bosons helicities. This requires the detection of the forward muons and the measurement of their azimuthal angular difference. Our strategy is similar to the one that is employed by the LHC experiments, see e.g.~Refs.~\cite{ATLAS:2022fnp,CMS:2022uox}, using the forward jets from VBF production~\cite{Hankele:2006ma}. We find that the forward muon detector would enable continuing these studies at a muon collider, with better sensitivity than the projections for the High-Luminosity LHC (HL-LHC) and future $e^+e^-$ colliders.

These findings rely on assumptions on the yet-unknown performances of the forward muon detector. In particular, the resolution on the measurement of the muon energies has an important impact on most (though not all) of our studies. We consider $10\%$ as baseline resolution, and discuss in the Conclusions how the physics potential would be reduced by an order $100\%$ resolution, which corresponds to not measuring the muon energies at all. The forward detector would still enable some of our physics studies by measuring the muon angles.

The rest of the paper is organised as follows. In Section~\ref{sec:general} we summarise our setup~\cite{Ruhdorfer:2023uea} for the simulation of the response of the forward muon detector and of the effects associated with the imperfections of the incoming muon beams. We also describe the simulation of background processes that are common to several of the analyses presented in the paper. In Section~\ref{sec:inclusive} we revisit the inclusive Higgs cross section and the invisible Higgs branching ratio sensitivity projections, updating existing results with improved background simulation and selection strategies. Section~\ref{sec:BSM} is devoted to the search of invisible BSM particles through the Higgs portal. In Section~\ref{sec:hZZ_CP} we explain how the measurement of the forward muons' azimuthal angle ``resurrects''~\cite{Panico:2017frx} the interference between vector bosons of different helicities in the initial state. We then illustrate the advantages of this interference resurrection mechanism for the study of the $CP$ properties of the $hZZ$ coupling.
Our conclusions are reported in Section~\ref{sec:conc}.

\section{Simulation setup and common backgrounds
}\label{sec:general}

We consider a forward muon detector with coverage on the muon pseudo-rapidity $\eta_\mu$ in the range
\begin{equation}
\eta_{\rm MD} < |\eta_\mu| < \eta_{\mu}^{\rm max},
\end{equation}
where $\eta_{\rm MD}$ is the acceptance of the Main general-purpose Detector. If not specified otherwise, we set $\eta_{\rm MD} = 2.44$ ($\theta_{\rm MD} = 10^{\rm o}$). We limit the coverage of the forward detector to $\eta_\mu^{\rm max} = 6$ ($\theta_\mu^{\rm min} = 0.005\,$rad), because this is enough to collect nearly all muons that are emitted in association with effective $Z$ bosons~\cite{Ruhdorfer:2023uea}. 

The trajectory of the forward muons crosses the conical absorbers and other elements of the collider ring. Therefore, it will be possible to detect them only if their energy is high enough to penetrate a long layer of dense material. Based on a preliminary assessment~\cite{private}, we set the lower energy threshold for muon detection to 500~GeV. We assume that the muons within the energy acceptance
\begin{equation}
    E_{\mu^\pm} > 500\;\mathrm{GeV}\,,
\end{equation}
are seen with $100\%$ probability, and the others are lost.

We also assume that the forward detector can measure the kinematic properties of the muons. The resolution of the muon energy measurement is simulated by a Gaussian smearing with constant relative uncertainty $\delta_{\rm res}$, and $\delta_{\rm res} = 10\%$ is used as benchmark value. The uncertainty on the muon direction is neglected in the benchmark setup.\footnote{Our baseline matches the simulation of the forward detector implemented in the muon collider \texttt{DELPHES} card~\cite{delphes_card_mucol}. The card assumes a $10\%$ smearing on the transverse  momentum rather than on the energy, which however is equivalent to the energy smearing in the absence of angular smearing. The card also includes a constant $95\%$ efficiency for muon detection. Our results do not include this efficiency if not specified otherwise.} Variations of the baseline parameters values will be discussed when relevant.

Monte Carlo (MC) parton-level data samples are generated with \texttt{MadGraph5\_aMC@NLO}~\cite{Alwall:2014hca}. The simulations account for the beam energy spread (BES) $\delta E/E = \delta_{\rm BES}$, which we set to $0.1\%$ in accordance with the muon collider target parameters~\cite{Accettura:2023ked}. The BES has two effects. First, the centre of mass energy of the initial muons, $\sqrt{s}$, is smeared around the nominal value of $2E_{\rm b}=10$~TeV. Second, the centre of mass frame of the collision is boosted along the beam axis, with respect to the detector frame, owing to the unbalance in the muons' energy. 

The simulation of the BES is not automated in \texttt{MadGraph}. We thus include the BES effects by proceeding as follows~\cite{Ruhdorfer:2023uea}.  Truth-level $\mu^+\mu^-$ collision events are generated in the centre of mass frame, for different values of $\sqrt{s}$. The $\sqrt{s}$ distribution---which is approximately Gaussian with mean $2E_{\rm{b}}$ and standard deviation $\sigma=\sqrt{2}\, \delta_{\rm BES}E_{\rm{b}}$---is sampled at the three values of $\{2E_{\rm{b}} -\sigma, 2E_{\rm{b}}, 2E_{\rm{b}} +\sigma\}$. These three datasets are eventually combined with equal weights of $1/3$. The rapidity of the centre of mass frame---whose distribution conditional to $\sqrt{s}$ is approximately Gaussian with zero mean and standard deviation $\delta_{\rm BES}/(2\sqrt{2})$---is introduced by reprocessing the three samples. For each truth-level event, the boost rapidity is sampled from its distribution and the corresponding Lorentz transformation is applied to the final-state particles. The procedure was validated against the BES implementation that is available in \texttt{WHIZARD}~\cite{Moretti:2001zz,Kilian:2007gr}. 

The angular divergence of the beam at the interaction point produces a Beam Angular Spread (BAS) that could be included in the simulation with a similar strategy~\cite{Ruhdorfer:2023uea}. The BAS is neglected in the benchmark setup because its effects are small.

In the studies considered in this paper, the signal is characterised by two forward energetic opposite-charge muons emitted in opposite hemispheres. Event preselection thus requires the observation of one $\mu^+$ and one $\mu^-$, subject to acceptance cuts
\begin{equation}\label{eq:acceptance_cuts}
\begin{split}
|&\eta_{\mu}| < 6\,,\qquad \;\,E_{\mu^\pm} > 500\;\mathrm{GeV},  \\ \eta_{\mu^+}&\cdot\, \eta_{\mu^-} < 0\,,\qquad \Delta R_{\mu\mu} > 0.4\,.
\end{split}
\end{equation}

The upper limit on the absolute rapidity ensures that the muons will be observed, either by the main detector or by the forward detector. A lower limit that selects forward muons could be imposed, but it is typically not needed since the background muons are as forward as the ones from the signal. 

The other angular cuts select muons in opposite hemispheres and with enough separation---using the variable $\Delta{R}=\sqrt{(\Delta\eta)^2+(\Delta\phi)^2}$---to eliminate the contribution from the decay of $Z$ bosons or virtual photons emitted orthogonally to the beam line. 

The elastic (Bhabha) scattering \mbox{$\mu^+\mu^- \to \mu^+\mu^-$} produces forward muons and it is a priori a very large background for those signals in which the forward muons are the only visible or tagged particles in the final state. We suppress Bhabha scattering by a lower cut on the total transverse momentum of the muon pair, which we take to be $P_\perp^{\mu\mu} > 50\;\mathrm{GeV}$ or larger in our studies. The cut has a moderate impact on the VBS or VBF signal topologies, where the forward muons have typical transverse momentum of order $m_Z$. After the cut, the relevant backgrounds are processes that produce the forward energetic muons with significant total momentum in the transverse plane. The rest of the section is devoted to the description of these processes. 
\begin{figure*}[t!]
\centering
\includegraphics[width=0.48\textwidth]{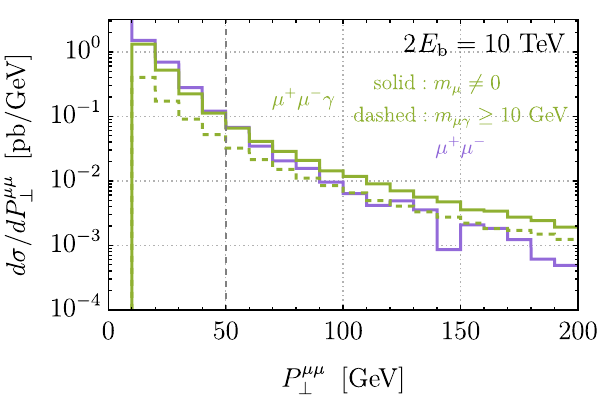} \hfill
\includegraphics[width=0.48\textwidth]{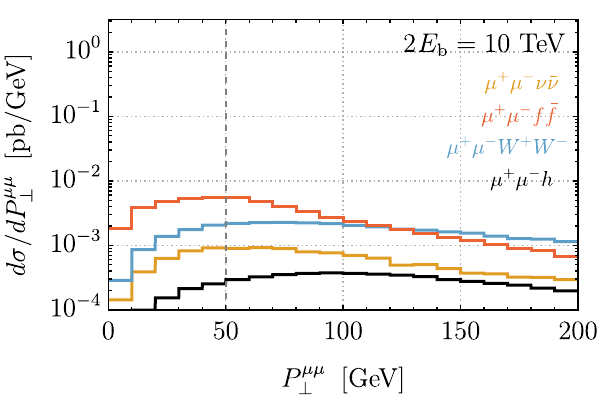}
\caption{\label{fig:photonBGpTmumu}{\bf Left:} Truth-level distribution of $P_\perp^{\mu\mu} = (p_{\mu^+} + p_{\mu^-})_\perp$ using different descriptions of the $\mu^+ \mu^-$ plus photons process. Only the acceptance cuts in Eq.~\eqref{eq:acceptance_cuts} are applied. The dashed vertical line indicates the (minimum) cut applied in our baseline selection. {\bf Right:} The distributions for other backgrounds and for an example signal, namely Higgs production in $ZZ$ fusion.}
\end{figure*}

The first background is the Bhabha process accompanied by radiation of photons that generate some amount of $P_\perp^{\mu\mu}$. 
This background cannot be suppressed by a photon veto, because the photons are typically soft or collinear either with the incoming or with the outgoing forward muons. Therefore, they are outside the (energy or angular) acceptance of the main detector.

The most accurate strategy to simulate the process would be to generate a merged sample of $\mu^+ \mu^- \to \mu^+\mu^-$ plus additional photons matched to QED showering. However, matching techniques are automated for QCD, while QED matching is not implemented in any of the available multi-purpose MC generators and we need to consider alternative descriptions of the process. 

One possibility is to generate elastic $\mu^+\mu^-$ scattering followed by~\texttt{PYTHIA8}~\cite{Bierlich:2022pfr} photon showering. Notice that the proper inclusion of initial state radiation effects, based on backwards evolution, requires that the muon parton distribution function (PDF) is employed in the fixed-order event generation. As the muon PDF is not directly available in \texttt{MadGraph}, this was achieved using instead the electron PDF implementation~\cite{Frixione:2021zdp} but replacing the mass of the electron with the one of the muon.

Care is also needed in the choice of the cutoff scale for the showering. The default \texttt{PYTHIA8} settings employ the centre of mass energy~$\sqrt{s}$, while the energy transferred in the scattering is best estimated by the transverse momentum of the muons. We thus pick the square root of (minus) the Mandelstam variable $t$ as the scale for \texttt{PYTHIA8} showering. In the relevant kinematic regime with forward muons, $\sqrt{-t}$ is of order tens or at most hundreds of~GeV and much smaller than the default scale $\sqrt{s}=10$~TeV, producing much softer radiation.

With this simulation method, based on parton showering, it is computationally expensive to populate the $P_\perp^{\mu\mu} > 50$~GeV region. Furthermore, the usage of a showering description is fully justified only in the presence of a large separation between the scale of the radiation and the one of the hard process. There is not really a separation for such relatively large values of $P_\perp^{\mu\mu}$, which are on the contrary comparable with the hard scale $\sqrt{-t}$.

An alternative description is offered by the tree-level process $\mu^+ \mu^- \to \mu^+\mu^-\gamma$, which we generate with a $p_\perp^\gamma =P_\perp^{\mu\mu} > 10$~GeV lower cut. In order to regulate the residual singularity associated with collinear photon emission from final-state muons, we retain the finite mass of the muon in the \texttt{MadGraph} event generation. We checked the stability of the result by varying the muon mass around its physical value and observing a successful event generation and a smooth dependence of the cross section on the mass. This suggests that \texttt{MadGraph} could successfully achieve the integration over the phase space.\footnote{In its current version, \texttt{MadGraph} event generation consistently fails if instead the muon mass is set to zero and the singularity is not regulated. However, older versions such as \texttt{v3.5.1} produce events even for $m_\mu = 0$. Using this un-physical simulation we recovered the $\mu \mu\gamma$ background used in Ref.~\cite{Li:2024joa}.
\label{footnote2}}

The left panel of Fig.~\ref{fig:photonBGpTmumu} shows the $P_\perp^{\mu\mu}$ distributions of the $\mu \mu \gamma$ fixed-order sample (solid green) and the $\mu \mu$ showered sample (solid violet). A relatively good agreement is observed for $P_\perp^{\mu\mu} \approx 50$~GeV, but the fixed-order prediction features a more pronounced high-$P_\perp^{\mu\mu}$ tail. Since we consider fixed-order more reliable, and we aim at conservative results, in what follows we use the $\mu \mu\gamma$ simulation to describe Bhabha plus photons. 

The plot also shows (dashed green) the distribution obtained from a $\mu \mu\gamma$ sample generated with vanishing muon mass, but with a cut $m_{\mu\gamma} > 10$~GeV that regulates the collinear singularity. The latter simulation, which we employed in Ref.~\cite{Ruhdorfer:2023uea}, underestimates the cross section by an $\mathcal{O}(1)$ factor and will not be used in this work.

The $\mu\mu\gamma$ background---and, to some extent, the comparison with the showered prediction---is quite sensitive even to a moderate lower cut on the transverse momentum of individual muons, $p_{\perp}^{\mu}$.
This is because the photon is radiated from one of the muons, which receives a significant transverse kick, while the other muon remains very forward and with a transverse momentum typically below $\mathcal{O}(10)$~GeV. The emission of multiple photons mitigates the unbalance in the transverse momentum of the muons. Therefore, for a better modelling of the $p_{\perp}^{\mu}$ distribution 
we also include in our background a simulation of the two-photons $\mu \mu\gamma\gamma$  production. It will turn out that this process plays a negligible role in all our analyses, after the full selection cuts are applied.

Other backgrounds are $\mu^+ \mu^- \nu \bar{\nu}$ and $\mu^+ \mu^- f\bar{f}$, where $f$ denotes any quark or charged lepton. We treat these two processes separately because only the latter can be partially vetoed when appropriate---see later---by exploiting the main detector. The $\mu\mu f\bar{f}$ process includes the production of a virtual photon decaying to $f\bar{f}$. We eliminate the corresponding singularity by applying a $10$~GeV cut on the invariant mass of the $f\bar{f}$ pair. The region below the cut corresponds to the splitting of a real photon to $f\bar{f}$ and we consider that it should be accounted for in a sufficiently realistic way by the $\mu\mu \gamma$ simulation. 

We also include the $\mu^+\mu^- W^+ W^-$ background, which we simulate including the $W$ boson decays to fermion pairs using~\texttt{MadSpin}~\cite{Artoisenet:2012st}, unlike in Ref.~\cite{Ruhdorfer:2023uea} where the $W$ bosons were not decayed.

The right panel of Fig.~\ref{fig:photonBGpTmumu} shows the $P_\perp^{\mu\mu}$ distributions for these backgrounds, together with the example signal $\mu^+ \mu^- \to \mu^+ \mu^- h$. Their contribution in the region defined by the acceptance cuts of Eq.~\eqref{eq:acceptance_cuts} is much smaller than the Bhabha background reported in the left panel. However, they will have a considerable impact after the final selection cuts in most of the analyses of the paper.

\begin{figure*}[t]
\centering
\includegraphics[width=0.48\textwidth]{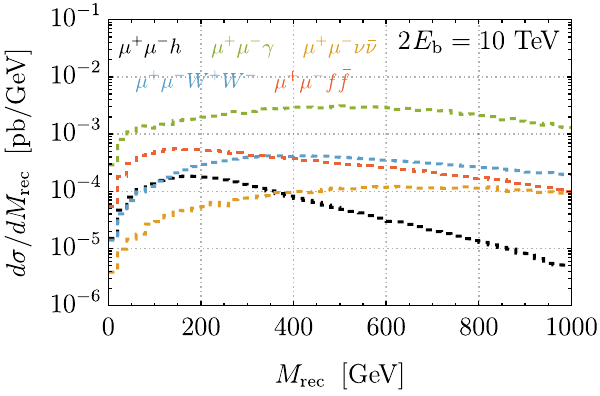} \hfill
\includegraphics[width=0.48\textwidth]
{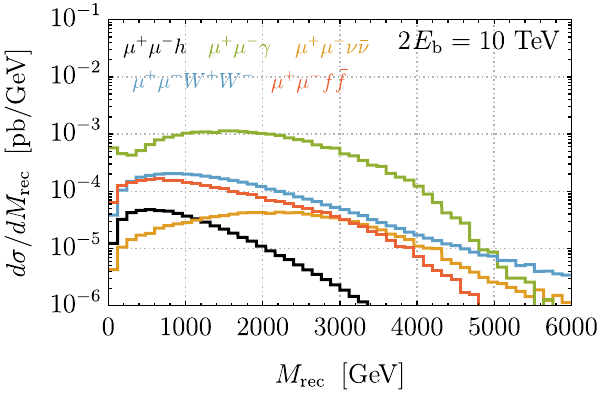}
\caption{\label{fig:incHiggsMrec}{\bf Left:}
Signal and background $M_{\rm rec}$ distributions after the baseline cuts for the inclusive Higgs analysis described in Section~\ref{sec:ihp}. A muon energy resolution $\delta_{\rm res} = 1\%$ is assumed. {\bf Right:} The same, but assuming $\delta_{\rm res} = 10\%$. Note the different scale on the horizontal axis.}
\end{figure*}

The statistical methodology employed in each analysis is detailed in the corresponding section. If not specified otherwise, our sensitivity or exclusion reach projections only include statistical uncertainties and neglect systematics and theory errors. It is worth emphasising that this is in general \emph{not} justified by the state-of-the-art knowledge of the muon collider experimental environment and the accuracy of the theoretical predictions. For instance, Higgs physics measurements have a statistical precision at the per mille level, and it is yet to be demonstrated that the corresponding theoretical predictions can reach a comparable or better level of accuracy. Similar considerations apply to the luminosity and other sources of experimental systematics. This cautionary remark applies to many muon collider sensitivity projection studies beyond the ones of the present paper. 

Our studies also pose specific challenges. At the theoretical level, they require an accurate prediction---or, alternatively, the experimental determination---of the SM processes that produce forward energetic muons, such as Bhabha scattering. Experimentally, they require an accurate calibration of the response of the forward muon detector. For instance, the resolution $\delta_{\rm res}$ of the muon energy measurement has a strong impact on the shape of the distributions that are relevant for our analyses. An accurate knowledge of this parameter will be necessary.

\section{Inclusive Higgs production and invisible Higgs decay}\label{sec:inclusive}

\begin{figure*}[t]
\centering
\includegraphics[width=0.48\textwidth]{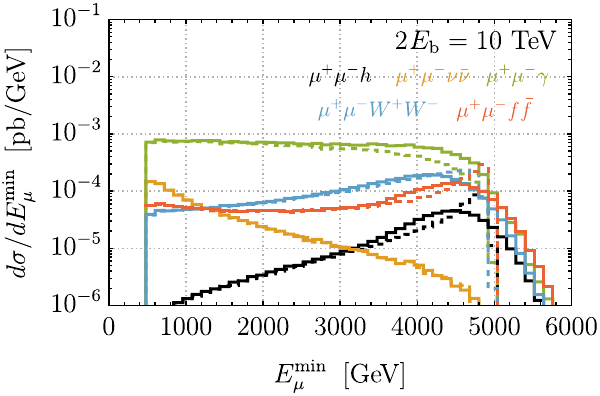} \hfill
\includegraphics[width=0.48\textwidth]{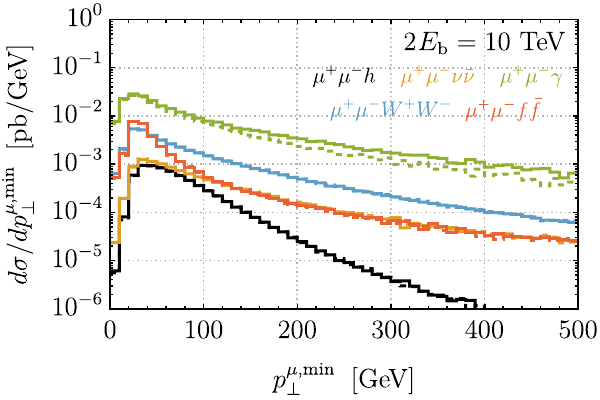}
\caption{\label{fig:incHiggsDist}{\bf Left:} Signal and background $E_{\mu}^{\rm min}$ distributions after the baseline cuts for the inclusive Higgs analysis described in Section~\ref{sec:ihp}. Dashed~(solid) lines assume a muon energy resolution $\delta_{\rm res} = 1\%\,(10\%)$. {\bf Right:} The same, but for $p_{\perp}^{\mu, \rm min}$.}
\end{figure*}

In this section we consider the on-shell production of the Higgs boson in $ZZ$ fusion
\begin{equation}\label{eq:h_ZZfusion}
\mu^+ \mu^- \rightarrow \mu^+ \mu^- h\,,
\end{equation}
aiming at an assessment of the sensitivity to the inclusive Higgs production cross section and to the invisible branching ratio of the Higgs. These studies are described in Sections~\ref{sec:ihp} and~\ref{eq:inv_Higgs} in turn.

\subsection{Inclusive Higgs production}\label{sec:ihp}

Our goal is to estimate the relative precision $\Delta \sigma/\sigma_{\rm SM}$ that can be achieved in the measurement of the cross section of the process in Eq.~\eqref{eq:h_ZZfusion}. We will assume a central value equal to the SM prediction $\sigma_{\rm SM}=87.4$~fb at the $2 E_{\rm b} = 10$~TeV muon collider. 

For the measurement we select events with two muons of opposite charge  satisfying the acceptance cuts in Eq.~\eqref{eq:acceptance_cuts}, allowing for the possible presence of any other particle in the final state in addition to the muons. This is because we aim at a determination of the cross section that is independent of the Higgs branching ratio in any standard or exotic decay channel. With this inclusive selection, all the processes described in Section~\ref{sec:general} contribute to the background. In order to mitigate the Bhabha background we apply a $P_\perp^{\mu\mu} > 50\;\mathrm{GeV}$ transverse momentum imbalance cut as previously explained. When considering the $\mu\mu f\bar{f}$ processes, one needs to subtract the contribution from the Higgs boson decay to fermions that is part of the signal and not a  background.

On the event samples defined by the baseline cuts (i.e., Eq.~\eqref{eq:acceptance_cuts} and $P_\perp^{\mu\mu} > 50\;\mathrm{GeV}$) we identify several kinematic variables that help separating the Higgs signal from the background in the final selection. Namely, we consider the invariant mass of the system recoiling against the final-state muons (recoil mass)
\begin{equation}\label{eq:recoil_mass}
M_{\rm rec} = \sqrt{|(\Delta P)^2|}\,, \quad \Delta P =  (2E_{\rm b}, \vec{0}\,) - p_{\mu^+} - p_{\mu^-} \,,
\end{equation}
as well as the minimal muon energy and transverse momentum
\begin{equation}
E_\mu^{\rm min}=\mathrm{min}{\{E_{\mu^+}},\,{E_{\mu^-}}
\}\,,\; p_{\perp}^{\mu,\rm min}=\mathrm{min}{\{p_\bot^{\mu^+}},\,{p_\bot^{\mu^-}}
\}\,.
\end{equation}
The signal and background distributions of these variables are shown in Figs.~\ref{fig:incHiggsMrec} and~\ref{fig:incHiggsDist} under two different assumptions on the resolution of the muon energy measurement achieved by the forward muon detector. We consider our benchmark resolution $\delta_{\rm res} = 10\%$ and a better (possibly unrealistic) resolution $\delta_{\rm res} = 1\%$.

The $M_{\rm rec}$ distribution and its discriminating power is strongly affected by the muon energy resolution. The Higgs mass peak is washed out already for $\delta_{\rm res} = 1\%$, but the signal distribution is still localised below around 400~GeV. This provides a handle for the rejection of some of the backgrounds, which extend to higher masses like the left panel of Fig.~\ref{fig:incHiggsMrec} shows. With $\delta_{\rm res} = 1\%$, a good sensitivity to the cross section is obtained with the cuts
\begin{equation}\label{eq:cuts_smear_1p}
\begin{split}
M_{\rm rec} <  \,&400\,\mathrm{GeV}\,,\quad E_\mu^{\rm min} > 4\,\mathrm{TeV}\,,\\
&\;p_{\perp}^{\mu,\rm min} < 200\,\mathrm{GeV}\,.
\end{split}
\end{equation}
The upper panel of Table~\ref{tab:results_inclusive_1percent} shows the number of signal and background expected events with $10\;\mathrm{ab}^{-1}$ luminosity after the baseline cuts and the final selection in Eq.~(\ref{eq:cuts_smear_1p}). The corresponding $68\%$ CL relative uncertainty on the cross section measurement is $(\Delta \sigma / \sigma_{\rm SM})_{68\%}= 0.38\%$.

By inspecting the cut-flow, we observed that the $E_\mu^{\rm min}$ cut is mostly efficient to suppress the $\mu\mu\gamma$ and $\mu\mu \nu\bar{\nu}$ backgrounds, whereas the impact of the $p_{\perp}^{\mu,\rm min}$ cut is similar for all backgrounds. This is compatible with the shapes of the $E_\mu^{\rm min}$ and $p_{\perp}^{\mu,\rm min}$ distributions displayed in Fig.~\ref{fig:incHiggsDist}. The cut on $M_{\rm rec}$ is found to have a subdominant effect. By excluding this cut from Eq.~\eqref{eq:cuts_smear_1p} we obtain a relatively mild degradation of the sensitivity to $(\Delta \sigma / \sigma_{\rm SM})_{68\%}= 0.41\%$.

With the benchmark energy resolution, $\delta_{\rm res} = 10\%$, the recoil mass loses discriminating power, as Fig.~\ref{fig:incHiggsMrec} shows. The most effective selection strategy in this case is to eliminate the $M_{\rm rec}$ cut and to relax the minimal muon energy requirement in comparison with Eq.~\eqref{eq:cuts_smear_1p}, namely
\begin{equation} \label{eq:cuts_smear}
E_\mu^{\rm min} > 3.5\,\mathrm{TeV}\,,\qquad p_{\perp}^{\mu,\rm min} < 200\,\mathrm{GeV}\,.
\end{equation}
Using these selections we estimate---see the lower panel of Table~\ref{tab:results_inclusive_1percent}---a sensitivity $(\Delta \sigma / \sigma_{\rm SM})_{68\%}= 0.50\%$, with a still reasonably high signal-over-background ratio of $8\%$.

\renewcommand{\tabcolsep}{6pt}
\renewcommand{\arraystretch}{1.4}
\begin{table}[h]
\centering
\begin{tabular}{ lccc }    
$\delta_{\rm{res}}=1\%$  & baseline cuts & final selection, Eq.~\eqref{eq:cuts_smear_1p}   \\\hline
$\mu^+\mu^- h$  & $6.8 \cdot 10^5$ & $4.3 \cdot 10^5$      \\
$\mu^+\mu^- \gamma$ & $2.4 \cdot 10^7$ & $3.4 \cdot 10^5$     \\
$\mu^+\mu^- \nu \bar{\nu}$ & $1.3 \cdot 10^6$ & $1.0 \cdot 10^4$    \\
$\mu^+\mu^- f \bar{f}$ & $3.3\cdot 10^6$ & $1.2\cdot 10^6$    \\
$\mu^+\mu^- W^+ W^-$ & $4.6\cdot 10^6$ & $6.0\cdot 10^5$   \\\hline
$(\Delta \sigma / \sigma_{\rm SM})_{68\%}$ & & $0.38\%$    \\
$S_{\rm SM}/B$ & & $0.20$  \vspace{1.5mm}
\end{tabular}
\centering
\begin{tabular}{ lccc } 
$\delta_{\rm{res}}=10\%$\hfil  & baseline cuts  &  final selection, Eq.~\eqref{eq:cuts_smear} \\\hline
$\mu^+\mu^- h$  & $6.8 \cdot 10^5$ & $5.3 \cdot 10^5$   \\
$\mu^+\mu^- \gamma$ & $2.9 \cdot 10^7$ & $2.7 \cdot 10^6$  \\
$\mu^+\mu^- \nu \bar{\nu}$ & $1.3\cdot 10^6$ & $3.6\cdot 10^4$   \\
$\mu^+\mu^- f \bar{f}$ & $3.3 \cdot 10^6$ & $1.7 \cdot 10^6$ \\
$\mu^+\mu^- W^+ W^-$ & $4.6\cdot 10^6$ & $1.9\cdot 10^6$ \\\hline
$(\Delta \sigma / \sigma_{\rm SM})_{68\%}$ &  & $0.50\%$ \\
$S_{\rm SM}/B$ &  & $0.08$
\end{tabular}
\caption{{\bf{Top:}} Number of events expected for the inclusive Higgs cross section measurement, with  $10\,\mathrm{ab}^{-1}$, assuming $\delta_{\rm res} = 1\%$. The $\mu^+\mu^- h$  cross section is set to the SM value. The expected precision is estimated as $(\Delta \sigma / \sigma_{\rm SM})_{68\%} = \sqrt{B + S_{\rm SM}}/S_{\rm SM}$.
{{\bf{Bottom:}} The same, but for $\delta_{\rm res} = 10\%$.}
}
\label{tab:results_inclusive_1percent}
\end{table}


In summary, our projections for the inclusive Higgs production cross section sensitivity are
\begin{equation}
    ( \Delta \sigma  / \sigma_{\rm SM} )^{ZZ\rightarrow h}_{68\%} = \begin{cases}
        0.38\%\quad \, \,\,\delta_{\rm res} = 1\%\\
        0.50\% \quad\, \,\,\delta_{\rm res} = 10\%
    \end{cases}.
\end{equation}
Our result is better than the one ($0.75\%$, for $\delta_{\rm res} = 10\%$) reported in Ref.~\cite{Li:2024joa}, where the inclusive Higgs measurement was first studied. This is due to an overestimated $\mu \mu \gamma$ background (see Footnote~\ref{footnote2}) and suboptimal cuts. 

The inclusive cross section measurement can be turned into a $hZZ$ coupling determination that is independent of the other Higgs couplings and of the Higgs decay width. With the benchmark detector configuration $\delta_{\rm res} = 10\%$, this measurement corresponds to a constraint $|\delta g_{hZZ}/g_{hZZ}^{\rm SM}| < 2.5 \cdot 10^{-3}$ at $68\%$~CL on the coupling deviation from the SM value. This is very close to the projected sensitivity of FCC-ee of $1.7\cdot 10^{-3}$~\cite{deBlas:2019rxi}.

\subsection{Invisible Higgs decay}\label{eq:inv_Higgs}

We turn now to the determination of the Higgs to invisible decay branching ratio. The results that follow are an update of our previous studies presented in Ref.~\cite{Ruhdorfer:2023uea}.

The baseline event selection requires the acceptance cuts~\eqref{eq:acceptance_cuts} and $P_\perp^{\mu\mu} > 50$~GeV like in the previous section. In addition, since we now target the invisible decay of the Higgs, we apply a veto on any other visible object (photon, jet, or charged lepton) in the main detector. 

The effectiveness of the veto on the visible particles depends on the acceptance and the reconstruction efficiency of the main detector. As in Ref.~\cite{Ruhdorfer:2023uea}, we assume an angular acceptance $|\eta| < \eta_{\rm{MD}} = 2.44$ and a transverse momentum threshold $p_\perp > 20$~GeV for the observation of any visible object. The results reported below assume $100\%$ reconstruction efficiency, but we verified that a mis-reconstruction rate at the few per mille level would not degrade the performances.

Final selection cuts are identified~\cite{Ruhdorfer:2023uea} to enhance the sensitivity to the signal. For $\delta_{\rm res} = 1\%$, these are
\begin{equation}
\begin{split}
&|\Delta \eta_{\mu\mu}| > 8\,,\; |\Delta \phi_{\mu\mu} - \pi| > 0.8\,,\; P_\perp^{\mu\mu} > 80\;\mathrm{GeV}, \\
&M_{\mu\mu} > 9.5\,\mathrm{TeV},\; E_\mu^{\rm min} > 4.7\,\mathrm{TeV},\; M_{\rm rec} < 0.8\,\mathrm{TeV},
\end{split}
\end{equation}
where the $M_{\rm rec}$ variable defined in Eq.~\eqref{eq:recoil_mass} coincides with the Missing Invariant Mass (MIM) used in Ref.~\cite{Ruhdorfer:2023uea}. 

For $\delta_{\rm res} = 10\%$, $M_{\rm rec}$ is no longer a useful discriminating variable and the best sensitivity is obtained with the selection cuts
\begin{equation}
\begin{split}
|\Delta \eta_{\mu\mu}| >&\; 6.5\,,\;\, |\Delta \phi_{\mu\mu} - \pi| > 1\,,\;\, P_\perp^{\mu\mu} > 180\;\mathrm{GeV}\,, \\
&M_{\mu\mu} > 8.75\;\mathrm{TeV}\,,\quad E_\mu^{\rm min} > 4.3\;\mathrm{TeV}\,.
\end{split}
\end{equation}
We remark, in preparation for the discussion of Section~\ref{sec:BSM}, that the most important variable to separate the signal from the backgrounds is $E_\mu^{\rm min}$. The $M_{\mu\mu}$ variable also plays an important role. 

The only updates of our analysis in comparison with the one of Ref.~\cite{Ruhdorfer:2023uea} concern the simulation of the following two background processes:
\begin{itemize}[leftmargin=10pt]
\item 
In the $\mu\mu W W$ sample the $W$'s are decayed to two-particle final states using~\texttt{MadSpin}, and the main detector veto is applied on the decay products unlike in Ref.~\cite{Ruhdorfer:2023uea}, where the veto was applied directly on the undecayed $W$'s. Decaying the $W$'s increases the background cross section, because of the following. The $W$ can decay leptonically in a configuration where most of its $p_\bot$ is carried by the invisible neutrino---which cannot be vetoed---while the charged lepton transverse momentum is below the detection threshold. This makes the $W$ invisible even if its transverse momentum and pseudo-rapidity are inside the main detector acceptance. By inspecting the $\mu\mu W W$ background events after the selection cuts ($75\%$ of which come from the leptonic decay of both $W$'s) we verified that the dominant topology is indeed the one with central and high-$p_\bot$ neutrinos, while the charged leptons are just below the $p_\bot$ threshold for the main detector veto.
\item 
As discussed in Section~\ref{sec:general}, the collinear singularity of the $\mu \mu \gamma$ background process is regulated by the physical muon mass, instead of the $m_{\mu \gamma} > 10$~GeV cut. This leads to a moderate increase of the cross section.
\end{itemize}

With these new background simulations, in the benchmark scenarios with a forward detector acceptance of $|\eta_\mu|<6$, we obtain the $95\%$~CL bounds\footnote{The sensitivity to the invisible Higgs branching ratio is below the SM value BR$_{\rm inv}^{\rm SM} = 1.2 \cdot 10^{-3}$. We thus express our results, like in Ref.~\cite{Ruhdorfer:2023uea}, as a limit on the additional BSM contribution BR$_{\rm inv}^{\rm BSM}$ under the hypothesis that the branching ratio is as predicted by the SM. Since BR$_{\rm inv}^{\rm BSM}$ is positive, we set the limits as one-sided exclusions.}
\begin{equation}
{\rm{BR}}_{\rm inv}^{\rm BSM} <
     \begin{cases}
     4.6 \cdot 10^{-4}\quad \, \,\,\,\,\delta_{\rm res} = 1\%\\
        1.3 \cdot 10^{-3} \quad\, \,\,\,\,\delta_{\rm res} = 10\%
    \end{cases}.
\end{equation}
These are approximately $10\%$ weaker than the constraints quoted in Ref.~\cite{Ruhdorfer:2023uea}, as a result of the mild background increases discussed above.

Reference~\cite{Forslund:2023reu} recently provided another estimate of the invisible branching ratio sensitivity. For $\delta_{\rm res}=10\%$ and $|\eta_\mu| < 6$, the estimated sensitivity is approximately a factor of two weaker than ours: BR$_{\rm inv} < 2.2\cdot 10^{-3}$. We verified that most of the discrepancy stems from a less optimal choice of the selection cuts. With the event selection of Ref.~\cite{Forslund:2023reu}, namely
\begin{equation}
\begin{split}\label{eq:cuts_meade}
    p_\bot^{\mu, {\rm min}} > &\;20\text{ GeV}\,,\;\; P_\bot^{\mu\mu} > 100\text{ GeV}\,,\;\; \Delta R_{\mu \mu} > 9\,, \\
    &M_{\mu\mu} > 9\text{ TeV}\,,\quad 2.5 < |\eta_{\mu}| < 6\,,
\end{split}
\end{equation}
we obtain indeed BR$_{\rm inv} < 2.6\cdot 10^{-3}$, which is in fair agreement with Ref.~\cite{Forslund:2023reu}, though slightly weaker. The residual discrepancy is due to the fact that the muon PDF was not used in Ref.~\cite{Forslund:2023reu} for the signal simulation, while we include it in our simulation of the signals involving invisible final states. The signal cross section after the cuts in Eq.~\eqref{eq:cuts_meade} is approximately $20\%$ higher if the muon PDF is not employed and the result of Ref.~\cite{Forslund:2023reu} is recovered.

\section{Invisible scalars through off-shell Higgs}\label{sec:BSM}

Here we turn to the pair-production of a BSM particle $\phi$ mediated by an off-shell Higgs boson, namely
\begin{equation}\label{eq:h_ZZfusion_phiphi}
\mu^+ \mu^- \rightarrow \mu^+ \mu^- (h^\ast \to \phi\phi)\,.
\end{equation}
We assume that $\phi$---which we take to be a real scalar for definiteness---is an invisible particle and has a mass $m_\phi > m_h/2$. Note that if instead $m_\phi < m_h /2$, the $\phi$ particle would contribute to the invisible Higgs branching ratio and could be probed using the results of Section~\ref{eq:inv_Higgs}.

\begin{figure*}[t]
\centering
\includegraphics[width=0.48\textwidth]{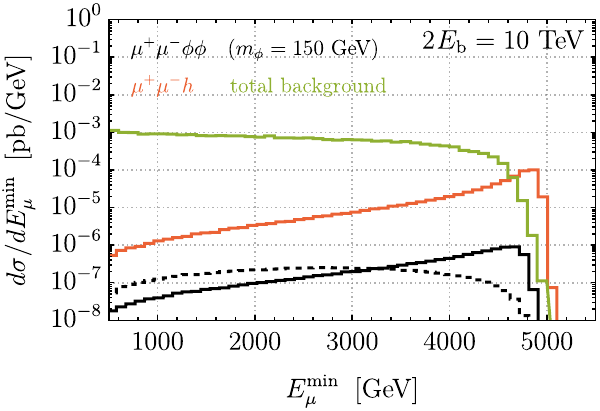} \hfill
\includegraphics[width=0.48\textwidth]{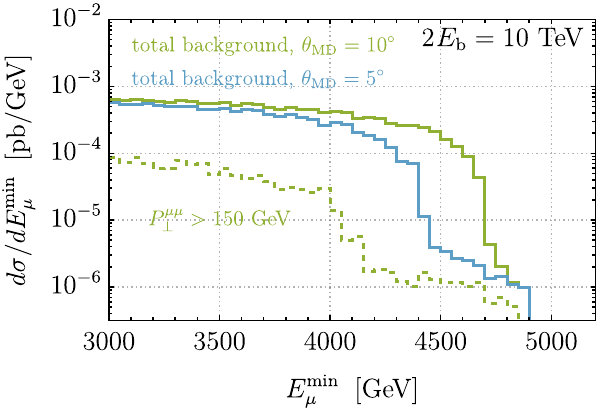}
\caption{\label{fig:EmuMinPortal} Distributions of $E_{\mu}^{\rm min}$ for the invisible Higgs portal signals and main backgrounds after baseline cuts, assuming a muon energy resolution $\delta_{\rm res} = 1\%$. {\bf Left:} The solid~(dashed) black histogram corresponds to the renormalisable (derivative) Higgs portal signal for $\lambda=4$~($f = 500$~GeV) and $m_\phi = 150$~GeV. The on-shell Higgs signal is shown in red for comparison, while the combined backgrounds are shown in green. {\bf Right:} Combined background distributions after baseline cuts for the benchmark main detector configuration $\theta_{\rm MD} = 10^{\rm o}$ (solid green) and an extended main detector coverage $\theta_{\rm MD} = 5^{\rm o}$ (light blue). The dashed green histogram assumes benchmark main detector coverage, but enforces an additional $P_\bot^{\mu\mu}>150$~GeV cut.}
\end{figure*}

We study two distinct scenarios for the portal interaction between $\phi$ and the SM Higgs doublet $H$. The first one is a renormalisable coupling~\cite{Silveira:1985rk,McDonald:1993ex,Burgess:2000yq}, with BSM Lagrangian
\begin{equation}\label{eq:RenormPortal}
    \mathcal{L}_{\rm ren} = \frac{1}{2}(\partial_\mu \phi)^2 - \frac{1}{2}M_\phi^2 \phi^2 -\, \frac{\lambda}{2}\phi^2 H^\dagger H\,.
\end{equation}
As a model of thermal dark matter with standard cosmological history, this scenario has been excluded by direct searches for Weakly-Interacting Massive Particles (WIMPs)~\cite{LZ:2024zvo} for $m_\phi \lesssim 30$~TeV (with the exception of a narrow region around $m_\phi \sim m_h/2$). Here $m_\phi = (M_\phi^2 + \lambda v^2/2)^{1/2}$ is the physical mass of the scalar, with $v \approx 246\;\mathrm{GeV}$ the Higgs vacuum expectation value. Nevertheless, non-standard cosmological histories can open up broad swaths of parameter space~\cite{Hardy:2018bph}. In addition, the signature we study can apply to scenarios where $\phi$ does not contribute appreciably to the dark matter abundance, but plays a role in addressing other outstanding questions, such as in models of electroweak baryogenesis~\cite{Curtin:2014jma} or neutral naturalness~\cite{Cohen:2018mgv,Cheng:2018gvu}. In these scenarios $\phi$ is not necessarily stable, but may decay invisibly or be sufficiently long lived to escape the detector. For instance, two complex scalar top partners transforming in the fundamental representation of an $SU(3)$ symmetry and coupled to the Higgs field via $\mathcal{L} \supset -\, y_t^2  (|\tilde{u}_1^c|^2 + |\tilde{u}_2^c|^2) H^\dagger H$ provide a solution to the Higgs naturalness problem. If the top partners are invisible, their production contributes to the signature considered here. Our results based on Eq.~\eqref{eq:RenormPortal} can be directly applied to the case where the top partners are mass-degenerate, by simply accounting for the relevant multiplicities: the $\tilde{u}_i^c$ are equivalent to a single real scalar $\phi$ with effective coupling strength $\lambda = \sqrt{12}\, y_t^2$. Hence, this value of the coupling provides an interesting theoretical target.

The second case we consider is a derivative coupling
\begin{equation}\label{eq:DerivativePortal}
    \mathcal{L}_{\rm der} = \frac{1}{2}(\partial_\mu \phi)^2 - \frac{1}{2}m_\phi^2 \phi^2 + \frac{1}{2f^2}\,\partial_\mu (\phi^2)\, \partial^\mu (H^\dagger H)\,.
\end{equation}
This is motivated for example by composite Higgs models where both $\phi$ and $H$ arise as pseudo-Goldstone bosons of a spontaneously broken global symmetry~\cite{Frigerio:2012uc,Balkin:2018tma}. Provided $\phi$ is stabilised by the symmetries of the model, it can be a WIMP dark matter candidate with $s$-wave annihilation to SM particles mediated by the operator in Eq.~\eqref{eq:DerivativePortal}. The same operator leads to momentum-suppressed scattering on nuclei, easily escaping the bounds from direct dark matter searches. Constraints from indirect detection observatories are rather weak, with Fermi-LAT~\cite{Fermi-LAT:2016uux} only excluding DM masses just above the $m_\phi = m_h/2$ threshold~\cite{Ruhdorfer:2019utl}. Hence, the vast majority of the parameter region where $\phi$ provides the observed dark matter abundance via standard thermal freeze-out is currently untested. For $m_\phi \gtrsim 130$~GeV this region is described by the simple relation $f \simeq 1.3\;\mathrm{TeV}\,(m_\phi/130\;\mathrm{GeV})^{1/2}$.

\subsection{Signal vs background discrimination}

The analysis proceeds along similar lines as for the study of the invisible Higgs decay in Section~\ref{eq:inv_Higgs}. The baseline selection requires the acceptance cuts~\eqref{eq:acceptance_cuts}, $P_\perp^{\mu\mu}> 50$~GeV and the main detector veto. However, further background rejection turns out to be more difficult, and increasingly challenging as $m_\phi$ is raised above the $m_h/2$ threshold. 

The variables with highest discriminating power in the invisible Higgs analysis---primarily $E_\mu^{\rm min}$, but $M_{\mu\mu}$ also plays an important role---are less useful. This occurs because the forward muons lose more energy in order to produce a final state---the $\phi\phi$ pair---that has higher invariant mass than the Higgs mass since $2\hspace{0.2mm}m_\phi>m_h$. Therefore, the signal $E_\mu^{\rm min}$ distribution is shifted to lower values and is more contaminated by the background. Similar considerations hold for the variable $M_{\mu\mu}$, which also becomes less discriminant. The $E_\mu^{\rm min}$ distribution, in comparison with the background and invisible Higgs signal distributions, is shown in the left panel of Fig.~\ref{fig:EmuMinPortal} for $m_\phi = 150$~GeV and representative values of the portal couplings in Eqs.~\eqref{eq:RenormPortal} and~\eqref{eq:DerivativePortal}.

In addition to a lower threshold, the Higgs portal signal distributions also feature a smooth threshold behaviour that further reduces the discriminating power in comparison with the invisible Higgs, whose distribution is instead sharply peaked close to the threshold. The effect is due to the dependence on the invariant mass of the cross section of the underlying vector boson scattering process, $ZZ\to\phi\phi$. In the case of the renormalisable coupling, the cross section has a maximum close to the $2\hspace{0.2mm}m_\phi$ threshold, but extends to larger invariant mass. This produces a smoother $E_\mu^{\rm min}$ distribution (solid black line) compared to the case of the Higgs signal (red line), where the underlying fusion process $ZZ\to h$ is strongly peaked at the Higgs mass. In the case of the derivative portal coupling, the effective interaction strength increases with energy. This pushes the $ZZ\to\phi\phi$ cross section to higher invariant mass and produces an $E_\mu^{\rm min}$ distribution that is much broader (dashed black line) than the one of the renormalisable portal model. 

The endpoint of the $E_\mu^{\rm min}$ distribution for the background processes is controlled  mainly by two parameters: the $P_\bot^{\mu\mu}$ cut and the angular coverage of the main detector that is assumed in the veto. The right panel of Fig.~\ref{fig:EmuMinPortal} shows that lowering the background distribution endpoint---which is potentially helpful for signal selection---can be accomplished by either imposing a stronger cut $P_\bot^{\mu\mu}>150$~GeV, or by assuming an extended angular coverage $\theta_{\rm MD} = 5^{\rm o}$ instead of the benchmark coverage $\theta_{\rm MD} = 10^{\rm o}$ (i.e., $\eta_{\rm MD}=2.44$).

\renewcommand{\tabcolsep}{6pt}
\renewcommand{\arraystretch}{1.4}
\begin{table*}[p]
\centering
\begin{tabular}{ lccccc }    
\hfil [number of events, $10$~ab$^{-1}$] & $\mu^+ \mu^- \phi\phi$ signal &  $\mu^+\mu^-\bar{\nu}\nu$ &  $\mu^+\mu^- \gamma$ & $\mu^+\mu^- \bar{f} f$ & $\mu^+\mu^- W^+ W^-$ \\\hline
baseline  & $1.1\cdot 10^{2}\cdot \lambda^2$ & $1.1\cdot 10^{6}$  & $2.3\cdot 10^{7}$ & $1.5\cdot 10^{6}$ &  $3.7\cdot 10^{5}$  \\
$M_{\rm rec} > 0.6$~TeV & $97\cdot \lambda^2$ & $7.3\cdot 10^{5}$   & $1.1\cdot 10^{7}$ & $6.9\cdot 10^{5}$ & $3.1\cdot 10^{5}$  \\
$|\Delta \eta_{\mu\mu}| > 6$ & $96\cdot \lambda^2$ & $5.5\cdot 10^{5}$   & $1.0\cdot 10^{7}$ & $6.4\cdot 10^{5}$ & $2.8\cdot 10^{5}$  \\
$|\Delta\phi_{\mu\mu} - \pi| > 1$ & $71\cdot \lambda^2$ & $2.8\cdot 10^{5}$   & $1.6\cdot 10^{6}$ & $3.6\cdot 10^{5}$ & $1.6\cdot 10^{5}$ \\
$P_{\bot}^{\mu\mu} > 130$~GeV & $47\cdot \lambda^2$ & $1.4\cdot 10^{5}$ & $3.5\cdot 10^{5}$ & $1.1\cdot 10^{5}$ &  $7.9\cdot 10^{4}$  \\
$M_{\mu\mu} > 7.4$~TeV & $36\cdot \lambda^2$ & $7.7\cdot 10^{3}$ & $3.2\cdot 10^{4}$ & $1.5\cdot 10^{4}$ & $1.2\cdot 10^4$   \\
$E_{\mu}^{\rm min} > 4.1$~TeV & $15\cdot \lambda^2$ & $4.6\cdot 10^{2}$ & $42$ & $2.8\cdot 10^2$ &  $1.1\cdot 10^3$ 
\\\hline
\end{tabular}\caption{Cut-flow for $2E_{\rm b} = 10$~TeV, $\theta_{\rm MD} = 10^\circ$ and $\delta_{\rm res}=1\%$, for a scalar with mass $m_\phi = 300$~GeV interacting through the renormalisable Higgs portal. The cuts give a $95\%$ CL bound $\lambda < 2.2$, about $15\%$ worse than we obtain with the classifier.
}
\label{tab:cut_flow_1p_renorm}
\end{table*}

\renewcommand{\tabcolsep}{6pt}
\renewcommand{\arraystretch}{1.4}
\begin{table*}[p]
\centering
\begin{tabular}{ lccccc }    
\hfil [number of events, $10$~ab$^{-1}$] & $\mu^+ \mu^- \phi\phi$ signal &  $\mu^+\mu^-\bar{\nu}\nu$ &  $\mu^+\mu^- \gamma$ & $\mu^+\mu^- \bar{f} f$ & $\mu^+\mu^- W^+ W^-$ \\\hline
baseline  & $4.6\cdot 10^{2}\cdot \left(\text{TeV}/f\right)^4$ & $1.1\cdot 10^{6}$  & $2.3\cdot 10^{7}$ & $1.5\cdot 10^{6}$ &  $3.7\cdot 10^{5}$  \\
$M_{\rm rec} > 1.2$~TeV & $4.2\cdot 10^2\cdot \left(\text{TeV}/f\right)^4$ & $1.8\cdot 10^{5}$   & $6.8\cdot 10^{5}$ & $5.4\cdot 10^{4}$ & $2.4\cdot 10^{5}$  \\
$|\Delta \eta_{\mu\mu}| > 5.5$ & $4.2\cdot 10^2\cdot \left(\text{TeV}/f\right)^4$ & $1.4\cdot 10^{5}$   & $6.6\cdot 10^{5}$ & $5.1\cdot 10^{5}$ & $2.2\cdot 10^{5}$  \\
$|\Delta\phi_{\mu\mu} - \pi| > 1$ & $3.1\cdot 10^2\cdot \left(\text{TeV}/f\right)^4$ & $7.1\cdot 10^{4}$   & $1.3\cdot 10^{5}$ & $2.9\cdot 10^{4}$ & $1.3\cdot 10^{5}$ \\
$P_{\bot}^{\mu\mu} > 90$~GeV & $2.5\cdot 10^2\cdot \left(\text{TeV}/f\right)^4$ & $5.3\cdot 10^{4}$ & $4.9\cdot 10^{4}$ & $1.7\cdot 10^{4}$ &  $9.2\cdot 10^{4}$  \\
$M_{\mu\mu} > 3.5$~TeV & $2.4\cdot 10^2\cdot \left(\text{TeV}/f\right)^4$ & $3.5\cdot 10^{4}$ & $3.7\cdot 10^{4}$ & $1.4\cdot 10^{4}$ & $7.7\cdot 10^4$   \\
$E_{\mu}^{\rm min} > 1.8$~TeV & $1.9\cdot 10^2\cdot \left(\text{TeV}/f\right)^4$ & $8.0\cdot 10^{3}$ & $2.6\cdot 10^3$ & $4.2\cdot 10^3$ &  $5.3\cdot 10^4$ 
\\\hline
\end{tabular}\caption{Same as in Table~\ref{tab:cut_flow_1p_renorm}, but for the derivative Higgs portal. The chosen cuts give $f>790$~GeV at $95\%$ CL, about $16\%$ worse than the bound obtained with the classifier.}
\label{tab:cut_flow_1p_derivative}
\end{table*}

\renewcommand{\tabcolsep}{6pt}
\renewcommand{\arraystretch}{1.4}
\begin{table*}[p]
\centering
\begin{tabular}{ lccccc }    
\hfil [number of events, $10$~ab$^{-1}$] & $\mu^+ \mu^- \phi\phi$ signal &  $\mu^+\mu^-\bar{\nu}\nu$ &  $\mu^+\mu^- \gamma$ & $\mu^+\mu^- \bar{f} f$ & $\mu^+\mu^- W^+ W^-$ \\\hline
baseline  & $1.1\cdot 10^{2}\cdot \lambda^2$ & $1.1\cdot 10^{6}$  & $2.7\cdot 10^{7}$ & $1.5\cdot 10^{6}$ &  $3.7\cdot 10^{5}$  \\
$M_{\rm rec} > 0.6$~TeV & $91\cdot \lambda^2$ & $1.1\cdot 10^{6}$   & $2.5\cdot 10^{7}$ & $1.4\cdot 10^{6}$ & $3.5\cdot 10^{5}$  \\
$|\Delta \eta_{\mu\mu}| > 6$ & $90\cdot \lambda^2$ & $8.2\cdot 10^{5}$   & $2.3\cdot 10^{7}$ & $1.3\cdot 10^{6}$ & $3.3\cdot 10^{5}$  \\
$|\Delta\phi_{\mu\mu} - \pi| > 0.8$ & $73\cdot \lambda^2$ & $4.8\cdot 10^{5}$   & $3.5\cdot 10^{6}$ & $8.2\cdot 10^{5}$ & $2.1\cdot 10^{5}$ \\
$P_{\bot}^{\mu\mu} > 190$~GeV & $29\cdot \lambda^2$ & $1.3\cdot 10^{5}$ & $3.2\cdot 10^{5}$ & $9.5\cdot 10^{4}$ &  $5.1\cdot 10^{4}$  \\
$M_{\mu\mu} > 7.6$~TeV & $19\cdot \lambda^2$ & $1.2\cdot 10^{4}$ & $3.7\cdot 10^{4}$ & $1.2\cdot 10^{4}$ & $6.4\cdot 10^3$   \\
$E_{\mu}^{\rm min} > 3.9$~TeV & $9\cdot \lambda^2$ & $3.2\cdot 10^{3}$ & $1.6\cdot 10^3$ & $6.9\cdot 10^2$ &  $1.7\cdot 10^3$ 
\\\hline
\end{tabular}\caption{Cut-flow for $2E_{\rm b} = 10$~TeV, $\theta_{\rm MD} = 10^\circ$ and $\delta_{\rm res}=10\%$, for a scalar with mass $m_\phi = 300$~GeV interacting through the renormalisable Higgs portal. The cuts give a $95\%$ CL bound $\lambda < 3.9$, about $40\%$ worse than we obtain with the classifier.}
\label{tab:cut_flow_10p_renorm}
\end{table*}

\renewcommand{\tabcolsep}{6pt}
\renewcommand{\arraystretch}{1.4}
\begin{table*}[p]
\centering
\begin{tabular}{ lccccc }    
\hfil [number of events, $10$~ab$^{-1}$] & $\mu^+ \mu^- \phi\phi$ signal &  $\mu^+\mu^-\bar{\nu}\nu$ &  $\mu^+\mu^- \gamma$ & $\mu^+\mu^- \bar{f} f$ & $\mu^+\mu^- W^+ W^-$ \\\hline
baseline  & $4.6\cdot 10^{2}\cdot \left(\text{TeV}/f\right)^4$ & $1.1\cdot 10^{6}$  & $2.7\cdot 10^{7}$ & $1.5\cdot 10^{6}$ &  $3.7\cdot 10^{5}$  \\
$M_{\rm rec} > 3.5$~TeV & $1.8\cdot 10^2\cdot \left(\text{TeV}/f\right)^4$ & $1.5\cdot 10^{5}$   & $1.4\cdot 10^{6}$ & $8.5\cdot 10^{4}$ & $1.3\cdot 10^{5}$  \\
$|\Delta \eta_{\mu\mu}| > 5.5$ & $1.8\cdot 10^2\cdot \left(\text{TeV}/f\right)^4$ & $1.2\cdot 10^{5}$   & $1.4\cdot 10^{6}$ & $8.0\cdot 10^{4}$ & $1.2\cdot 10^{5}$  \\
$|\Delta\phi_{\mu\mu} - \pi| > 1$ & $1.4\cdot 10^2\cdot \left(\text{TeV}/f\right)^4$ & $8.2\cdot 10^{4}$   & $2.4\cdot 10^{5}$ & $4.2\cdot 10^{4}$ & $7.3\cdot 10^{4}$ \\
$P_{\bot}^{\mu\mu} > 90$~GeV & $1.0\cdot 10^2\cdot \left(\text{TeV}/f\right)^4$ & $4.6\cdot 10^{4}$ & $9.4\cdot 10^{4}$ & $2.3\cdot 10^{4}$ &  $5.3\cdot 10^{4}$  \\
$M_{\mu\mu} > 3.0$~TeV & $94\cdot \left(\text{TeV}/f\right)^4$ & $4.0\cdot 10^{4}$ & $9.0\cdot 10^{4}$ & $2.2\cdot 10^{4}$ & $4.4\cdot 10^4$   \\
$E_{\mu}^{\rm min} > 1.6$~TeV & $64\cdot \left(\text{TeV}/f\right)^4$ & $4.8\cdot 10^{3}$ & $1.4\cdot 10^4$ & $4.0\cdot 10^3$ &  $2.5\cdot 10^4$ 
\\\hline
\end{tabular}\caption{Same as in Table~\ref{tab:cut_flow_10p_renorm}, but for the derivative Higgs portal. The chosen cuts give $f> 640$~GeV at $95\%$ CL, about $35\%$ worse than the bound obtained with the classifier.}
\label{tab:cut_flow_10p_derivative}
\end{table*}

It should be noted that the background distribution close to the endpoint features a sharp drop, followed by a small nearly-flat tail. The tail is due to the $\mu\mu\nu\bar{\nu}$ background, and in fact it is insensitive to the main detector angular coverage as the neutrinos cannot be vetoed. 

The drop is due to the $\mu\mu X$ background processes, where $X = \gamma, f \bar{f}, W W$ is a visible object. The drop location can be understood as follows. The angle of $X$ can be estimated as $\theta_X \simeq p_\bot^X / p_z^X$, and the minimal transverse momentum $p_\bot^X=P_\bot^{\mu\mu}$ is dictated by the cut. The maximal longitudinal momentum $p_z^X$ can be estimated by the energy transferred by the muon to $X$, i.e. $E_{\rm b} - E_\mu^{\rm min}$. This sets a lower bound on $\theta_X$, which on the other hand needs to be smaller than $\theta_{\rm MD}$ for the event not to be vetoed. Hence 
\begin{equation}
E_\mu^{\rm min} \lesssim E_{\rm b} - \frac{P_\perp^{\mu\mu}}{\theta_{\rm MD}}\,.
\end{equation}
This estimate agrees well with the actual endpoints of the distributions shown in the right panel of Fig.~\ref{fig:EmuMinPortal}. For instance, assuming the benchmark main detector coverage $\theta_{\rm MD} = 10^{\rm o}$ and $P_\bot^{\mu\mu} > 50$~GeV~(150 GeV) yields estimates $E_\mu^{\rm min} \lesssim 4.7$~TeV (4.1 TeV). Based on this discussion, we expect a considerable improvement of the sensitivity with an extended $\theta_{\rm MD} = 5^{\rm o}$ angular coverage of the main detector. This is confirmed by the results presented in Section~\ref{sec:portal_results}.

We also investigated the possible advantages of a more aggressive $P_\bot^{\mu\mu}$ cut for a better separation of the invisible Higgs portal signals from the background. Still, a $P_\bot^{\mu\mu}$ cut larger than the baseline value of 50~GeV entails a significant reduction of the signal (notice that the $P_\bot^{\mu\mu}$ distribution is similar to the one of the on-shell Higgs signal shown in  Fig.~\ref{fig:photonBGpTmumu}). On the whole, the best sensitivity will emerge from selection cuts that are more inclusive than those for the invisible Higgs decay analysis. 

\begin{figure*}[t!]
\centering
\includegraphics[width=0.49\textwidth]{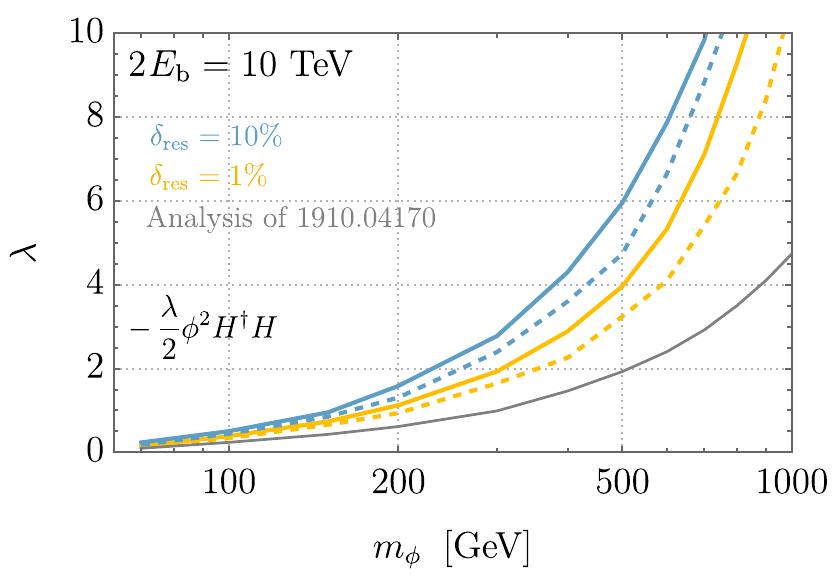}
\hfill
\includegraphics[width=0.49\textwidth]{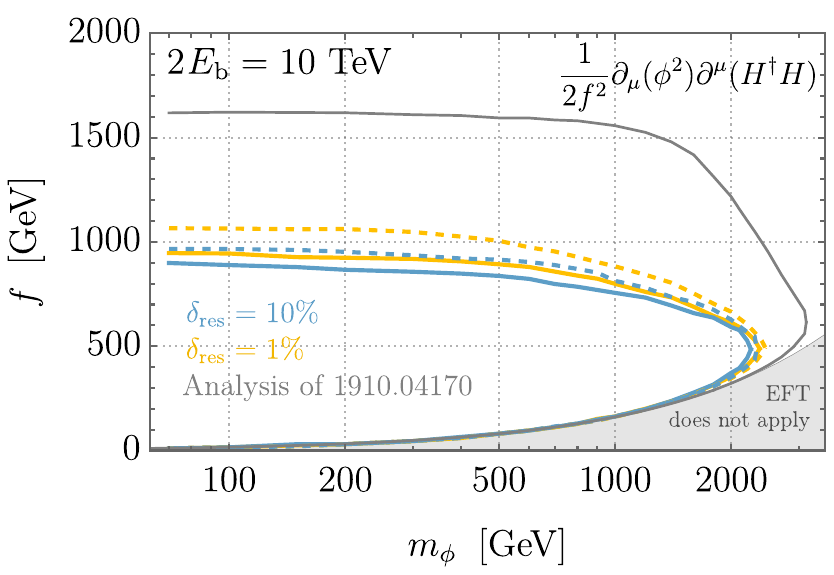}
\caption{\label{fig:margDetectorVariation} {\bf Left:} Projected $95\%$ CL bounds on the parameter space of the renormalisable Higgs portal model, derived from pair-production of invisible scalars at a $10$~TeV muon collider. Solid colored lines assume the benchmark angular coverage of the main detector $(10^\circ)$. Dashed colored lines assume an extended coverage $(5^\circ)$, which increases the effectiveness of the main detector veto. The thinner gray line shows, for reference, the unrealistically optimistic results of~\cite{Ruhdorfer:2019utl}. {\bf Right:} The same, but for the derivative Higgs portal model.}
\end{figure*}

Since the background rejection is very challenging, for an assessment of the optimal achievable sensitivity we will employ a feedforward neural network classifier.\footnote{The corresponding code is publicly available on \texttt{GitHub}~\cite{GitHub_link}.} We give 11 features of input to the classifier, namely
\begin{equation}
\begin{split}
        &\log \big(p_\bot^{\mu^{\pm}} \big)\,,\;\;\; \eta_{\mu^{\pm}}\,,\;\;\; \cos(\Delta\phi_{\mu\mu})\,,\;\;\;  \sin(\Delta\phi_{\mu\mu})\,, \\
       &|\Delta\eta_{\mu\mu}|\,,\quad M_{\rm rec}\,,\quad \log (P_\bot^{\mu\mu})\,,\quad M_{\mu\mu}\,,\quad E_\mu^{\rm min}\,,
\end{split}
\end{equation}
which provide a redundant description of the kinematics of the muons. Our projected constraints are computed  by requiring $S/\sqrt{B+S} = 1.64$ (corresponding to $95\%$~CL for a one-sided limit), both when employing the classifier and when applying manually-optimised cuts.

In Tables~\ref{tab:cut_flow_1p_renorm} and~\ref{tab:cut_flow_1p_derivative} we report optimised cut-flows for both portals, for a representative scalar mass $m_\phi = 300\;\mathrm{GeV}$, assuming a muon energy resolution $\delta_{\rm res} = 1\%$. The sensitivity we obtain with these manually-optimised cuts is quite close to the one we find with the classifier, which gives a $15\%$ ($30\%$) improvement of the projected bound on $\lambda$ ($1/f^2$). For our benchmark resolution $\delta_{\rm res} = 10\%$, the optimised cuts are reported in Tables~\ref{tab:cut_flow_10p_renorm} and~\ref{tab:cut_flow_10p_derivative}. In this case the manual cut optimisation is more challenging, because the worse muon energy resolution blurs the signal features in a more pronounced way. As a consequence, the classifier improves the performance significantly, allowing us to set constraints on $\lambda$ ($1/f^2$) that are roughly $40\%$ ($70\%$) better. In our final plots and results we use the bounds obtained with the classifier. 

\subsection{Results and comparison to other probes}\label{sec:portal_results}

The achievable bounds on the Higgs portal couplings are shown in Fig.~\ref{fig:margDetectorVariation}. We report results with variable muon energy resolution, $\delta_{\rm res} = 10\%$ or $1\%$, and angular coverage of the main detector $\theta_{\rm MD} = 10^\circ$ or $5^\circ$. For the renormalisable portal (left panel) the region above each line is excluded. For the derivative portal (right panel) the region inside each contour is excluded. We have shaded in gray the part of parameter space where the scale $f$ is too small for our description of the signal in terms of a dimension-$6$ Effective Field Theory (EFT) operator to be consistent.\footnote{As in Ref.~\cite{Ruhdorfer:2019utl}, we impose on the signal MC events an upper bound on the $\phi\phi$ invariant mass, $M_{\phi \phi} < g_\ast f$ with $g_\ast = 4\pi$. In the gray region, defined by $4\pi f < 2\hspace{0.2mm}m_\phi$, the bound can never be satisfied.}

We find that extending the main detector coverage from the current benchmark of $10^\circ$ down to $5^\circ$ would improve the sensitivity significantly, thanks to the increased effectiveness of the veto on $\mu\mu X$ background processes. The gain in sensitivity is comparable to the one expected if the muon energy resolution is improved from $10\%$ to $1\%$. We also find that our bounds are weaker than reported in Ref.~\cite{Ruhdorfer:2019utl}. In fact, in Ref.~\cite{Ruhdorfer:2019utl} the sensitivity was overestimated for two reasons. First, it was unrealistically assumed that all background processes except $\mu\mu \nu\bar{\nu}$ could be vetoed with perfect accuracy. Second, the effects of the BES and of the finite resolution of the forward muon detector were not considered and as a result the recoil mass was found to be a powerful discriminator between signal and background, which is instead not the case in our analysis.

\begin{figure*}[t!]
\centering
\includegraphics[width=0.49\textwidth]
{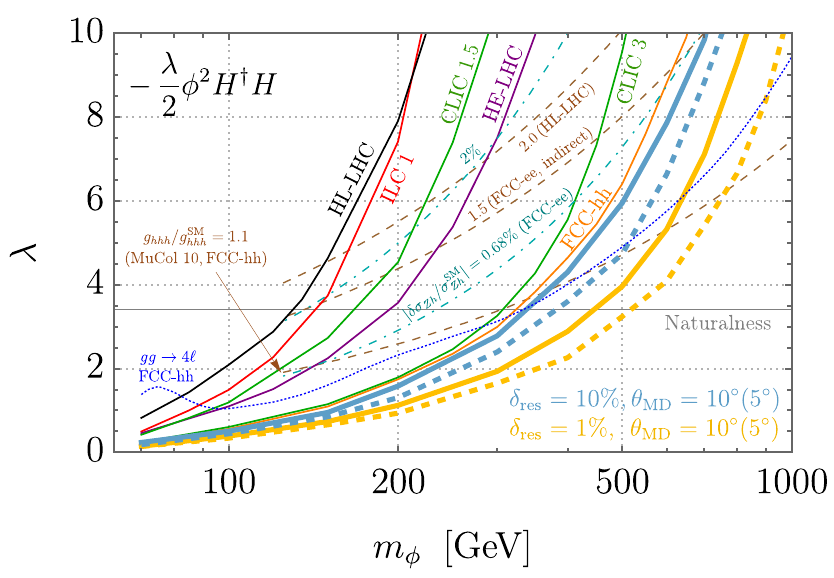}
\hfill
\includegraphics[width=0.49\textwidth]
{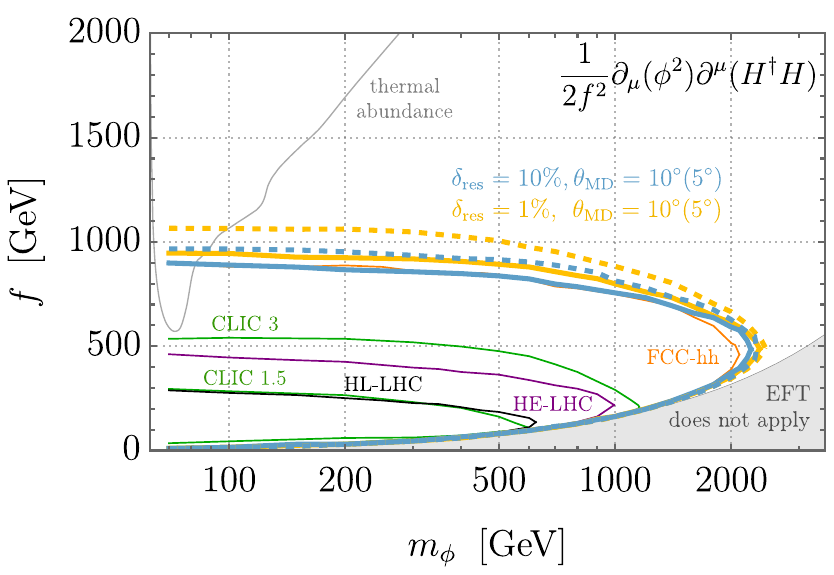}
\caption{\label{fig:portalComp} {\bf Left:} Parameter space of the renormalisable Higgs portal to invisible scalars $\phi$. The {\bf \color{deflightblue}thick light blue} and {\color{defgolden}\bf thick golden} lines show the exclusions we derived from $\phi$ pair-production in VBF at the 10 TeV muon collider, for different assumptions on the detector configuration. Thin solid lines show VBF bounds at hadron colliders (HL-LHC, {\color{defpurple}HE-LHC,} {\color{orange}FCC-hh)} and high-energy $e^+ e^-$ colliders {\color{red}(ILC,} {\color{defgreen}CLIC)}. {\color{defcyan}Thin dot-dashed cyan lines} show the one-loop correction to the $e^+ e^- \to Zh$ cross section. {\color{defbrown}Thin dashed brown lines} indicate the size of the one-loop correction to the $h^3$ coupling. The {\color{blue}thin dotted blue line} shows the FCC-hh sensitivity via the off-shell Higgs contribution to $gg\to ZZ\to 4 \ell$ production. All bounds are at $95\%$ CL. The {\color{gray}horizontal gray line} labeled ``Naturalness'' indicates the effective coupling strength corresponding to two mass-degenerate scalar top partners. {\bf Right:} The same, but for the derivative Higgs portal to invisible scalars. All contours show exclusions from $\phi$ pair-production in VBF at colliders, with {\bf \color{deflightblue}thick light blue} and {\color{defgolden}\bf thick golden} lines corresponding to our results for the 10 TeV muon collider. The {\color{deflightgray}thin gray contour} indicates the value of $f$ that produces via standard thermal freeze-out a present-day $\phi$ abundance matching the observed dark matter abundance. In the {\color{defdarkgray} dark gray-shaded region} a consistent EFT description of $\mu^+\mu^-\phi \phi$ production is not possible. See the main text (Section~\ref{sec:portal_results}) for further details.}
\end{figure*}

Qualitatively, the shapes of the bounds shown in Fig.~\ref{fig:margDetectorVariation} reflect the dependence on $m_\phi$ of the inclusive cross sections for $\mu^+ \mu^-\to  \mu^+ \mu^- \phi \phi$ production. For $m_h/2 \ll m_\phi \ll \sqrt{s}$ these are
\begin{equation}\label{eq:analyt_cross_sec}
\sigma_{\rm ren} \simeq  \frac{C_{ZZ} \lambda^2}{8 \pi m_\phi^2} \Big( \hspace{-0.75mm}\log \frac{s}{m_\phi^2} - \frac{14}{3} \Big)\,,\;\; \sigma_{\rm der} \simeq  \frac{3\hspace{0.25mm} C_{ZZ}  s}{16\pi f^4}\,,
\end{equation}
where $C_{ZZ} \equiv R_w^{\mu \bar{\mu}} g^4 / (6144\, \pi^4)$, $g$ is the SU(2)$_L$ gauge coupling and $R_w^{\mu \bar{\mu}} \approx 0.11$~\cite{Ruhdorfer:2019utl} accounts for the suppression of the muon couplings to the $Z$ relative to the $W^\pm$. Accordingly, for the renormalisable portal we observe that the bounds on $\lambda$ weaken rapidly as $m_\phi$ increases, whereas for the derivative portal the bounds on $f$ are nearly flat up to $m_\phi \sim 1$~TeV.

Lastly, in Fig.~\ref{fig:portalComp} we put our results into context by comparing them with other probes and outlining particularly motivated regions of the model parameter space.

We begin with the renormalisable portal (left panel of Fig.~\ref{fig:portalComp}). With our benchmark detector assumptions, $\delta_{\rm res} = 10\%$ and $\theta_{\rm MD} = 10^\circ$, we find that a 10~TeV muon collider matches the 100~TeV FCC-hh~\cite{Ruhdorfer:2019utl} (see also Ref.~\cite{Craig:2014lda}) in providing the strongest projected bounds. Extending the angular coverage of the main detector and/or improving the energy resolution of the forward muon detector leads to even better constraints, clearly surpassing FCC-hh. For example, if the portal coupling is fixed to the value required by naturalness of the Higgs mass ($\lambda = \sqrt{12}\, y_t^2$, horizontal gray line), our most optimistic estimate with $\delta_{\rm res} = 1\%$ and $\theta_{\rm MD} = 5^\circ$ reaches above $500$~GeV in the mass of the scalar top partners. For completeness, together with the muon collider and the FCC-hh we show the reach achievable in VBF production~\cite{Ruhdorfer:2019utl} at high-energy $e^+ e^-$ colliders (ILC at 1 TeV and CLIC at 1.5 and 3 TeV) and lower-energy proton colliders (HL-LHC at 14 TeV and HE-LHC at 27 TeV). 

Complementary probes of the renormalisable Higgs portal exploit one--loop effects. We consider three quantities for which a promising sensitivity has been demonstrated in the literature: the correction to the Higgs cubic coupling $g_{hhh}$~\cite{Curtin:2014jma,Englert:2019eyl}, the modification $\delta\sigma_{Zh}$ of the $e^+ e^- \to Zh$ cross section~\cite{Craig:2013xia}, and the off-shell Higgs contribution to $gg\to ZZ\to 4\ell$ production at hadron colliders~\cite{Haisch:2022rkm}. For the Higgs cubic coupling we show the contours $g_{hhh}/g_{hhh}^{\rm SM} = 2.0, 1.5$, and $1.1$ corresponding respectively to the approximate $95\%$ CL sensitivities of the HL-LHC, the FCC-ee, and the FCC-hh or a 10 TeV muon collider, which have comparable prospects~\cite{deBlas:2019rxi,Buttazzo:2020uzc}. For $e^+ e^- \to Zh$ we show the contour corresponding to the precision on $|\delta \sigma_{Zh}/\sigma_{Zh}^{\rm SM}|$ achievable at FCC-ee, $6.8 \cdot 10^{-3}$~\cite{deBlas:2019rxi}. We also draw the $2\cdot 10^{-2}$ precision contour for reference. Finally, for the off-shell Higgs contribution to $gg\to ZZ\to 4\ell$ we show the FCC-hh projection assuming $1\%$ systematic uncertainty~\cite{Haisch:2022rkm}.

We turn now to the derivative Higgs portal (right panel of Fig.~\ref{fig:portalComp}). Similarly to the case of the renormalisable portal, we find that a 10~TeV muon collider competes with FCC-hh for the best sensitivity. With benchmark detector assumptions $\delta_{\rm res} = 10\%$ and $\theta_{\rm MD} = 10^\circ$ we obtain a reach up to $f \sim 800$~GeV, extending above $1$~TeV for the most optimistic estimate. Focusing our attention on the scenario where $\phi$ is a thermal dark matter candidate, it appears that for $m_\phi \gtrsim 100$~GeV the corresponding parameter space is out of the reach of any of the considered future colliders. This motivates further studies going beyond the VBF topology, which could consider either the production of on-shell $\phi$ pairs via other processes, or loop corrections to precisely-measured observables. 

Concerning complementary probes, direct detection of WIMP dark matter interacting with the SM only by Eq.~\eqref{eq:DerivativePortal} will remain out of reach even in the future, since this operator mediates a negligibly small cross section for $\phi$ scattering on nuclei. By contrast, $\phi$ annihilation to SM particles is an $s$-wave process and the associated signal can be searched for with indirect detection experiments. Current constraints from Fermi-LAT~\cite{Fermi-LAT:2016uux} are weak, only excluding $m_\phi \lesssim 100$~GeV~\cite{Ruhdorfer:2019utl}, but improvements are expected in the next decade with the CTA observatory~\cite{CTA:2020qlo}.

\section{\boldmath{$CP$} violation in Higgs production}\label{sec:hZZ_CP}

In this section we describe and demonstrate the capability offered by the forward muon detector to measure the quantum-mechanical interference between the exchange of vector bosons with different helicity in VBS or VBF. The general strategy is described in Section~\ref{sec:IR}. Sections~\ref{sec:CPVHiggsTheory}, \ref{sec:CPVHiggsSignal} and 
\ref{sec:CPHTesting} consider an application to the study of the $CP$ property of the $hZZ$ coupling.

\subsection{Interference resurrection}\label{sec:IR}

We consider a generic scattering process
\begin{equation}\label{eq:VBSP}
    Z_1Z_2\to X\,,
\end{equation}
with $Z_{1,2}$ two $Z$ bosons moving along the positive and negative $z$-axis, respectively. The process can be observed at the muon collider through the emission of effective vector bosons~\cite{Fermi:1925fq,vonWeizsacker:1934nji,Dawson:1984gx,Kane:1984bb,Borel:2012by} from the incoming muons. Namely, we observe it in the reaction
\begin{equation}\label{eq:fullP}
    \mu^+\mu^-\to \mu^+\mu^- X\,,
\end{equation}
with forward muons in the final state. The first effective boson $Z_1$ in Eq.~(\ref{eq:VBSP}) is emitted from the incoming $\mu^+$ (which we take moving along the positive $z$-axis) in the $\mu^+\to \mu^+ Z_1$ splitting. The $Z_2$ boson is emitted in the splitting $\mu^-\to \mu^- Z_2$. The complete physical scattering process~(\ref{eq:fullP}) is well described by the hard collision of on-shell vector bosons~(\ref{eq:VBSP}), but only in the kinematic regime where the invariant mass of the $X$ system is much above the vector boson mass and the transverse momentum of the final state muons~\cite{Borel:2012by}.

Actually, the physical scattering~(\ref{eq:fullP}) can also receive contributions from the processes with initial state photons, $\gamma\gamma\to X$ and $Z\gamma\to X$. In what follows we assume for simplicity that such photon-induced processes are negligible. Otherwise, the formalism we present here should be generalised to account for the contribution of effective photons, including terms that stem from the interference between the $Z$ and the photon exchange. The possibility of accessing $Z/\gamma$ interference---for instance, in vector boson scattering processes where the photon contribution is large---deserves further studies and is left to future work.

Within the above-described approximations, the differential cross section of the process~(\ref{eq:fullP}) takes the form
\begin{equation}\label{eq:EVA} d\sigma=
    \sum_{h_{1,2},}
    \sum_{h_{1,2}^\prime}
    d\rho_{h_{1},h_{1}^\prime}^{Z_1}    d\rho_{h_{2},h_{2}^\prime} ^{Z_2}d\rho^{\rm{H}}_{h_1,h_2,h_1^\prime,h_2^\prime}
    \,.
\end{equation}
In the equation, we denote as $d\rho^{Z_{1,2}}$ the density matrix associated with the muon splitting that produces the effective $Z_{1,2}$ boson. We denote as $d\rho^{\rm{H}}$ the density matrix of the hard process, i.e. of the on-shell vectors scattering in Eq.~(\ref{eq:VBSP}). The meaning of these quantities is discussed in detail below.

Up to a phase-space factor, the hard density matrix is the square of the helicity amplitudes of the hard process but with different helicity indices $h_{1,2}$ and $h_{1,2}^\prime$ in the amplitude and in the conjugate amplitude,
\begin{equation}\label{eq:Hden}  d\rho^{\rm{H}}_{h_1,h_2,h_1^\prime,h_2^\prime}\hspace{-2pt}\propto\hspace{-2pt} 
{\cal{A}}(Z_1^{h_1}Z_2^{h_2}\hspace{-2pt}\to\hspace{-2pt} X)
{\cal{A}}^*(Z_1^{h_1^\prime}Z_2^{h_2^\prime}\hspace{-2pt}\to\hspace{-2pt} X)\,.
\end{equation}
Notice that in Eq.~\eqref{eq:EVA} we implicitly assume an un-polarised final state $X$. Correspondingly, a sum over the polarisations of the $X$ system is understood in the hard density matrix. 

A relevant generalisation of Eq.~(\ref{eq:EVA})---including also density matrices associated with the decay of polarised particles---could be given when the final state $X$ contains massive vectors or other unstable particles, and angular correlations are measured for their decay products. In this case, one can access the interference between amplitudes with particles of different helicity in the final state following Ref.~\cite{Panico:2017frx}. Here we target instead the interference between the helicities of the effective vector bosons in the initial state.

The $d\rho^{Z_{1,2}}$ density matrices in Eq.~(\ref{eq:EVA}) are proportional to the square of the  amplitudes that describe the splittings $\mu^+\to \mu^+ Z_1$ and $\mu^-\to \mu^- Z_2$, namely
\begin{equation}
\begin{split}
&d\rho_{h_{1},h_{1}^\prime}^{Z_{1}}\propto {\cal{A}}(\mu^+\to \mu^+ Z_1^{h_1})
    {\cal{A}}^*(\mu^+\to \mu^+ Z_1^{h_1^\prime})\,,\\
&d\rho_{h_{2},h_{2}^\prime}^{Z_{2}}\propto {\cal{A}}(\mu^-\to \mu^- Z_2^{h_2})
    {\cal{A}}^*(\mu^-\to \mu^- Z_2^{h_2^\prime})\,.
\end{split}
\end{equation}
An average over the helicities of the incoming (and outgoing) muons is understood in the previous equation, because we assume un-polarised muon beams. 

The analytic expression for the splitting amplitudes ${\mathcal{A}}$---see for instance Ref.~\cite{Nardi:2024tce}---is a function of the fraction $x$ of incoming muon (light-cone) momentum that is carried away by the effective $Z$ boson, of the norm of the transverse momentum of the final state muon, $p_\bot$, and of its azimuthal angle $\phi$. The distribution of these variables depends on the helicity of the vector boson, hence the measurement of any of them offers handles to isolate the contribution to the complete muon scattering process from the individual helicity amplitudes of the hard process. In what follows we focus on the azimuthal angle, which is particularly powerful for reasons that will become clear momentarily. 

The dependence on $\phi$ of the splitting amplitude is dictated by rotational symmetry. For a generic splitting process $A\to BC$, it reads
\begin{equation}
{\mathcal{A}}(A_{h_A}\to B_{h_B} C_{h_C} )\propto e^{-i(h_B+h_C-h_A)\phi}
\,.
\end{equation}
The expression further simplifies if $A$ and $B$ are massless muons, whose helicity is not changed by the chirality-preserving SM interactions. In this case, $h_B=h_A$ and the phase is proportional to the helicity $h_C$ of the emitted vector boson. The azimuthal angle dependence of the density  matrices is thus found to be
\begin{equation}\label{eq:denaz}    d\rho_{h_{1},h_{1}^\prime}^{Z_{1}}\propto e^{-i\,(h_1-h_1^\prime)\phi_+}\,,\quad
d\rho_{h_{2},h_{2}^\prime}^{Z_{2}}\propto e^{i\,(h_2-h_2^\prime)\phi_-}\,,
\end{equation}
where the sign flip in the $Z_2$ density matrix is due to the $Z_2$ boson moving along the negative $z$-axis. In the equation, $\phi_+$ and $\phi_-$ respectively denote the $\mu^+$ and $\mu^-$ azimuthal angles of spherical coordinates.

We finally remark that in the hard scattering process~(\ref{eq:VBSP}) the vectors can be considered to be exactly collinear with the beam line, as the recoil in the transverse plane against the muons emitted in the splitting is negligible. Therefore, the hard density matrix does not depend on the muon transverse momenta and in particular on the azimuthal angles. Each individual term of the helicity sums in the cross section formula~(\ref{eq:EVA})
\begin{equation}
{d\sigma}_{(h_1,h_2)\otimes(h_1^\prime,h_2^\prime)}=
    d\rho_{h_{1},h_{1}^\prime}^{Z_1}    d\rho_{h_{2},h_{2}^\prime} ^{Z_2}d\rho^{\rm{H}}_{h_1,h_2,h_1^\prime,h_2^\prime}\,,
\end{equation}
thus features a characteristic dependence on the $\phi_\pm$ (and $p_\bot$) variables---dictated by the dependence of the $Z_{1,2}$ density matrices---that makes it potentially observable. Using Eq.~(\ref{eq:denaz}), we find
\begin{equation}\label{eq:xsterms} {d\sigma}_{(h_1,h_2)\otimes(h_1^\prime,h_2^\prime)}\hspace{-2pt}\propto\hspace{-2pt}
    e^{-i\Delta h_1\phi_+}
    e^{i\Delta h_2\phi_-} d\rho^{\rm{H}}_{h_1,h_2,h_1^\prime,h_2^\prime}\,,
\end{equation}
where $\Delta h_{1,2}=h_{1,2}-h_{1,2}^\prime$.

The terms with $h_1=h_1^\prime$ and $h_2=h_2^\prime$ correspond to diagonal entries of the hard density matrix~(\ref{eq:Hden}). They are proportional to the modulus square of the individual helicity amplitudes and so in turn to the polarised $ZZ\to X$ scattering cross sections. The terms with $h_1\neq h_1^\prime$ and/or $h_2\neq h_2^\prime$ emerge from the quantum mechanical interference between the exchange of $Z$ bosons of different helicities and are proportional to the non-diagonal entries of the hard density matrix~(\ref{eq:Hden}). They depend on $\phi_\pm$ through oscillatory phases with a period $2\pi$ over the helicity differences $\Delta h_{1,2}$. Because of these phases, their contribution vanishes if the cross section is integrated over the azimuthal angles. The diagonal terms instead are $\phi_\pm$-$\,$independent and survive the integration.

In this context, the relevance of the forward muon detector stems from the fact that if the muons are not detected we only have access to observables that are inclusive (i.e., integrated) over the azimuthal angles. This makes interference effects unobservable because the angular integration kills their contribution. With the forward muon detector instead, we can measure the angles and ``resurrect'' the interference effects in suitably designed observables. A completely analogous mechanism was outlined in Ref.~\cite{Panico:2017frx} for the resurrection of interference effects associated with the helicity of unstable particles in the final state. In that case, the relevant azimuthal angles are those that characterise the unstable particle decays. Measuring those angles uniquely enables accessing the interference.

The possibility of exploiting angular correlations to access interference---both for initial- and final-state particles---has been known since long, see e.g. Refs.~\cite{Duncan:1985ij,Hagiwara:1986vm,Hankele:2006ma}. At the LHC, interference resurrection is starting to be deployed systematically~\cite{Panico:2017frx,Azatov:2019xxn,Hwang:2023wad,ElFaham:2024uop,Aoude:2023hxv,CMS:2021cxr,ATLAS:2019bsc,ATLAS:2022oge,ATLAS:2022vge} to enhance the sensitivity to EFT interaction operators, to access quantum mechanical entanglement, as well as for theory-agnostic measurements of the hard density matrix.

A muon collider would enable a similarly rich program, at higher energies and in the cleaner environment of muon collisions. As a simple illustration of this potential, in the rest of this section we show how the resurrection of the interference between the initial state vectors uniquely enables a precise test of the $CP$ properties of the $hZZ$ coupling.

\subsection{\boldmath{$CP$}-violating \boldmath{$hZZ$} coupling}\label{sec:CPVHiggsTheory}

Following Ref.~\cite{Azatov:2022kbs} we parametrise $CP$ violation by adding to the SM a $CP$-odd interaction, and the complete SM plus BSM $hZZ$ interaction Lagrangian reads
\begin{equation}\label{eq:hZZ}
    \mathcal{L}_{hZZ} =
    \frac{m_Z^2}{v}\, h\left[ c_{z}Z_\mu Z^\mu+
    \frac{\tilde{c}_{zz}}{v^2}
     \,Z_{\mu\nu} \widetilde{Z}^{\mu\nu}\right]\,,
\end{equation}
where $v \approx 246$~GeV, $m_Z$ is the $Z$ mass, $Z_{\mu\nu}$ and $\widetilde{Z}_{\mu\nu} \equiv  \epsilon_{\mu\nu\rho\sigma} Z^{\rho\sigma}/2$ are the field strength of the $Z$ boson and its dual. For reference, the strongest LHC constraint on the $CP$-odd parameter $\tilde{c}_{zz}$ is currently provided by CMS, $\tilde{c}_{zz}\in (-0.66, 0.51)$ at $95\%$~CL~\cite{CMS:2024bua}, with similar but slightly weaker limits obtained by ATLAS~\cite{ATLAS:2023mqy}.

In Eq.~(\ref{eq:hZZ}) we also introduced a parameter, $c_{z}$, that describes a $CP$-even deformation of the SM $hZZ$ coupling $c_{z}=1$. Even if our target is the $CP$-odd parameter $\tilde{c}_{zz}$, we retain $c_{z}\neq1$ in order to study the different impacts on the observables of the $CP$-even and $CP$-odd coupling components. This would be particularly relevant in the event of the observation of a non-SM Higgs coupling strength: we could attribute it to $\tilde{c}_{zz}$---and in turn discover $CP$-violation in Higgs couplings---only through observables that distinguish $\tilde{c}_{zz}\neq0$ from the effects of a non-SM $CP$-preserving coupling ${c}_{z}\neq1$.

A more general description of $CP$ violation in the Higgs coupling to vector bosons would also foresee---see Appendix~\ref{sec:CP_details}---the possibility of $CP$-odd couplings of the Higgs involving the photon. Here we focus on the $hZZ$ vertex, setting to zero the BSM corrections to the $h\gamma\gamma$ and $hZ\gamma$ vertices, but notice that photon couplings could also be probed by generalising the results of the previous section to include $\gamma$ exchange and $Z/\gamma$ interference.

We can now straightforwardly compute the helicity amplitudes of the $ZZ$ fusion process
\begin{equation}\label{eq:zzhfus}
    Z_1Z_2\to h\,,
\end{equation}
that produces the Higgs through the interactions~(\ref{eq:hZZ}). 

Before proceeding, it should be emphasised that the physical Higgs production process at the muon collider cannot be described quantitatively in the formalism of the effective collision of on-shell vector bosons. In order for that to be a good approximation, the invariant mass of the system produced in the scattering should be much larger than $m_Z$, while the Higgs is light. In fact, it is below the $2m_Z$ threshold so that the process~(\ref{eq:zzhfus}) cannot even occur with exactly on-shell $Z$ bosons. The analysis that follows---based on the effective $Z$ approximation---is thus intrinsically qualitative and aimed at identifying the relevant variables to access $\tilde{c}_{zz}$, rather than offering a quantitative description of Higgs production.

The helicity amplitudes read
\begin{equation}
\begin{split}
    &{\mathcal{A}}(Z_1^{h_1}Z_2^{h_2}\to h)=\\
    &\hspace{20pt}\frac{2\,m_Z^2}{v}\left[c_{z}\epsilon_{1,\mu}\epsilon_2^\mu
    -2\frac{\tilde{c}_{zz}}{v^2}
    \epsilon_{\mu\nu\rho\sigma}
    p_1^\mu \epsilon_1^\nu p_2^\rho \epsilon_2^\sigma
    \right]
    \,,
\end{split}
\end{equation}
where $\epsilon_{1,2}$ and $p_{1,2}$ denote the polarisation vectors and the momenta of the incoming $Z$ bosons. By substituting the explicit expressions\footnote{
We use: $\quad$$\epsilon_{1}^{h\,=\,\pm}=
\epsilon_{2}^{h\,=\,\mp}=
\{0,\mp\, 1/\sqrt{2},i/\sqrt{2},0\}$, \\
$p_{1(2)}=\{E,0,0,(-)p\}$, \ \ $\epsilon_{1(2)}^{h\,=\,0}=\{p/m_Z,0,0,(-)E/m_Z\}$.
}
 we get
 \begin{equation}
 \begin{split}\label{eq:HZZAMP}
    {\mathcal{A}}(Z_1^{\pm}Z_2^{\pm}\to h)
    =&\,\frac{2\,m_Z^2}{v}\left[
    c_{z}\mp 4\,i\,\frac{\tilde{c}_{zz}}{v^2}E\,p
    \right]\,,\\
    {\mathcal{A}}(Z_1^{0}Z_2^{0}\to h)
    =&\,\frac{2\,m_Z^2}{v}c_{z}\frac{E^2+p^2}{m_Z^2}\,,
    \end{split}
 \end{equation}
 while the other helicity amplitudes, with $h_1\neq h_2$, vanish by rotational symmetry since the Higgs is a scalar. In Eq.~\eqref{eq:HZZAMP}, $E$ and $p$ denote the energy and the momentum of the colliding $Z$ bosons in the centre of mass frame.

Thanks to the fact that only same-helicity hard amplitudes are non-vanishing, we can simplify the notation of the previous section by introducing a single index $h=h_1= h_2$ (and $h^\prime=h_1^\prime= h_2^\prime$) to label the different contributions to the cross section. Equation~\eqref{eq:xsterms} becomes
\begin{equation}\label{eq:termsHVBF}
d\sigma_{h\otimes h^\prime}\hspace{-2pt}\propto\hspace{-2pt}
    e^{-i\Delta h\,\Delta\phi}
 d\rho^{\rm{H}}_{h,h^\prime}\,,
\end{equation}
where $\Delta\phi = \phi_+-\phi_-$ and $\Delta h=h-h^\prime$. 

The helicity difference $\Delta{h}$ can assume the values $0$, $\pm1$ and $\pm2$. Therefore, owing to Eq.~(\ref{eq:termsHVBF}) the distribution of $\Delta\phi$ is the sum of a $\Delta\phi$-independent contribution, plus two $\Delta\phi$-dependent terms composed by trigonometric functions of $\Delta\phi$ and of $2\Delta\phi$, respectively. The constant term is the only one that does not vanish upon $\Delta\phi$ integration, hence it is proportional to the total cross section.

The constant term emerges from the diagonal entries of the density matrix, i.e. from the modulus square of the helicity amplitudes in Eq.~(\ref{eq:HZZAMP}). This term is not suited to probe $CP$ violation because the modulus square of the (transverse) amplitudes depends only the square of $\tilde{c}_{zz}$. In order to access small values of $\tilde{c}_{zz}$ we need instead linear effects, which are present only in the interference between different helicity amplitudes. 

Furthermore, the constant term receives contributions also from the $CP$-even coupling $c_{z}$. Therefore it does not discriminate a $CP$-odd from a $CP$-even component of the $hZZ$ interaction and would not allow us to establish the violation of $CP$ in the Higgs coupling if a departure from the SM was observed in the measurement. 

The terms that are proportional to $e^{i\,\Delta\phi}$ and to $e^{-i\,\Delta\phi}$ come from the interference between longitudinal and transverse $Z$ bosons, i.e. from $0\otimes+$ and $-\otimes0$ ($e^{i\,\Delta\phi}$), and from $0\otimes-$ and $+\otimes0$ ($e^{-i\,\Delta\phi}$). Their dependence on $\tilde{c}_{zz}$ is given by a $\tilde{c}_{zz}$-independent constant plus a linear term, since the transverse amplitudes are linear while the longitudinal amplitude is independent of $\tilde{c}_{zz}$. 

We finally have the  $+\otimes-$ and $-\otimes+$ interference terms. They are proportional to $e^{\mp2\,i\,\Delta\phi}$, respectively, and they contain a constant, a linear and a quadratic term in $\tilde{c}_{zz}$. 

The previous considerations restrict quite strongly the functional form of the $\Delta\phi$ distribution. Further constraints come from the $CP$ symmetry, by noticing that the $\Delta\phi$ variable is odd under $CP$ and that $\tilde{c}_{zz}$ is the only relevant parameter that breaks $CP$. By treating $\tilde{c}_{zz}$ as a spurion, we can thus conclude that the cross section must be invariant under the simultaneous change of sign of $\Delta\phi$ and $\tilde{c}_{zz}$. This selects the trigonometric function (sine or cosine) that appears in each term, leading to 
\begin{eqnarray}\label{eq:dspp}
\frac{2\pi}{\sigma_{h,\rm{SM}}}\frac{d\sigma_h}{d\Delta\phi}=&&
\;c_{z}^2 \left[1 +\kappa^c_{1}\cos(\Delta\phi)
+\kappa^c_{2}\cos(2\Delta\phi)\right]\nonumber\\
+&&
\;c_{z} \tilde{c}_{zz} \left[
\kappa_1^{l}\sin(\Delta\phi)+
\kappa_2^{l}\sin(2\Delta\phi)
\right]\\
+&&
\;\tilde{c}_{zz}^2
\left[
\kappa^q_0+
\kappa^q_2\cos(2\Delta\phi)
\right]\,.\nonumber
\end{eqnarray}
We will verify in Section~\ref{sec:CPHTesting} that this expression provides an excellent fit to the actual distribution and we will determine its coefficients.

In the rest of this section we estimate the muon collider sensitivity to $\tilde{c}_{zz}$. In order to probe small $\tilde{c}_{zz}$, we will need an observable that depends linearly and not quadratically on $\tilde{c}_{zz}$, i.e.~an observable that is sensitive to the terms on the second line of Eq.~(\ref{eq:dspp}). Since these terms vanish upon $\Delta\phi$ integration, the definition of such observable---concretely, an asymmetry---is based on the $\Delta\phi$ variable measured by the forward muon detector. A strong sensitivity improvement will be found in comparison with a strategy based on integrated observables such as the total cross section. Additionally, the simultaneous measurement of the asymmetry and of the total cross section enables an independent determination of $\tilde{c}_{zz}$ and of the $CP$-even coupling $c_{z}$. 

\subsection{Signal and background}\label{sec:CPVHiggsSignal}

Here we describe our simulation of the Higgs production process and of the backgrounds. We also identify the basic cuts that select Higgs production and reject the background, defining a suitable search region for the study of the $CP$-violating component of the Higgs production coupling that will be presented in Section~\ref{sec:CPHTesting}.

We consider production of the Higgs boson in $ZZ$ fusion followed by decay to $b\bar{b}$. The process of interest is 
\begin{equation}
    \mu^+\mu^- \rightarrow \mu^+ \mu^- (h\rightarrow b \bar{b})\,.
\end{equation}
For a sharp definition of the Higgs production process, we consider the Higgs to be effectively on-shell if the $b \bar{b}$ invariant mass is within $\pm15$ times the Higgs decay width around the Higgs mass. Backgrounds are the production of $\mu^+ \mu^- b \bar{b}$ with the $b \bar{b}$ invariant mass outside the Higgs window, $ \mu^+ \mu^- t \bar{t}$ where the $W$'s from the decay of the tops are not detected, as well as the $\mu^\pm \nu W^\mp h$ and $\mu \nu t b$ processes, which yield the $\mu^+\mu^- b\bar{b}\nu\bar{\nu}$ final state. The production of $\mu^+ \mu^- q \bar{q}$ with $q=u,d,c,s$, where the light quarks are misidentified as $b$-jets, is found to be negligible with the mis-identification probabilities (see below) that are assumed in the muon collider \texttt{DELPHES} card.

We generate MC events using \texttt{MadGraph} and a custom \texttt{FeynRules}~\cite{Alloul:2013bka} implementation of Eq.~\eqref{eq:hZZ}, which allows us to simulate the BSM effects due to the $\tilde{c}_{zz}$ and $c_z$ coefficients (we also cross-checked our implementation against the BSM Characterisation \texttt{FeynRules} model~\cite{BSM_characterization_FR}). We employ it in preparation for the $CP$ violation analysis of Section~\ref{sec:CPHTesting}, while the results of the present section are purely SM, i.e.~$\tilde{c}_{zz}=0$ and $c_{z}=1$.

The Higgs production sample is simulated as $\mu^+ \mu^- b \bar{b}$ production, with a cut enforcing the parton-level $b \bar{b}$ mass to be in the on-shell Higgs region defined previously.\footnote{Mild generation-level cuts are applied to the Higgs production and to the background simulations to avoid singularities, namely $|\eta_\mu| < 6.5$ and $\Delta R_{\mu\mu} > 0.4$. The finite $b$-quark mass is retained.} The normalisation is fixed by multiplying the on-shell $\mu^+ \mu^-\rightarrow \mu^+ \mu^- h$ tree-level cross section times the most precise branching ratio prediction~\cite{LHCHiggsCrossSectionWorkingGroup:2016ypw} BR$(h\rightarrow b\bar{b})_{\rm SM} = 0.582$. The simulation of the $\mu^+ \mu^- b \bar{b}$ background is performed with a cut on the $b \bar{b}$ mass that excludes the on-shell region. 

After tree-level generation, the events are passed to \texttt{PYTHIA8} for parton showering and next to \texttt{DELPHES}~\cite{deFavereau:2013fsa} for a simulation of the response of the muon collider detector~\cite{delphes_card_mucol} (with assumed main detector angular coverage of $\theta_{\rm MD} = 10^{\circ}$). Jet clustering is performed with \texttt{FastJet}~\cite{Cacciari:2011ma} using the Valencia algorithm~\cite{Boronat:2014hva}, with parameters set to $\beta=\gamma = 1$ and $R=0.5$ and requiring two jets in exclusive mode. Relevant parameters implemented in the muon collider \texttt{DELPHES} card are a constant $b$-tagging efficiency of $50\%$, and (energy- and rapidity-dependent) misidentification probabilities of less than $0.006$ for light jets and $\leq 0.03$ for $c\,$-jets. With these figures, the $\mu\mu q\bar{q}$ background is negligible as anticipated.

Unlike the other analyses described in the paper, the present one does not rely on the precise knowledge of the energy of the incoming muons. Therefore, the energy spread of the muon beams does not play a significant role and for this reason we do not implement it in our simulations. The present analysis is also nearly insensitive to the resolution in the measurement of the forward muon energies, because only the muon (azimuthal) angles will be exploited. A smearing on the muon energies at the $10\%$ level is nevertheless included in our simulations through the transverse momentum smearing implemented in the \texttt{DELPHES} card. A constant $95\%$ efficiency for muon detection is also included.

The forward muon detector is essential in order to observe Higgs production in $ZZ$ fusion, in particular to disentangle it from the dominant $WW$ fusion production mode. We use the information from the forward detector to require the presence of two forward muons subject to the acceptance cuts in Eq.~(\ref{eq:acceptance_cuts}).\footnote{In events with more than one $\mu^+$ and $\mu^-$ with energy larger than 500~GeV, we apply the angular cuts of Eq.~\eqref{eq:acceptance_cuts} to the $\mu^+$ and the $\mu^-$ with largest absolute rapidity in the appropriate hemisphere.} We also require, following Ref.~\cite{Forslund:2022xjq}, exactly $2$ $b$-jets satisfying
\begin{equation} \label{eq:cuts_bb}
    p_\bot^b > 30~\text{GeV},\qquad 
    M_{bb} \in [100, 150]~\text{GeV}\,.
\end{equation}
The corresponding event yields at $2 E_b = 10$~TeV and $L=10$~ab$^{-1}$ for the Higgs production process and for the backgrounds are reported in the first column of Table~\ref{tab:hbb_results}, assuming SM couplings for the Higgs. The cuts~(\ref{eq:cuts_bb}) achieve a good purity for the Higgs production signal, $S/B = 9.2$. 

We can use this result to estimate the expected statistical precision on the Higgs signal strength, obtaining $( \Delta \sigma  / \sigma^{\rm SM} )^{ZZ\rightarrow h\rightarrow b\bar{b}}_{68\%} = 0.75\%$. By repeating a similar study at the $2 E_b = 3$~TeV muon collider with 1~ab$^{-1}$ luminosity, we find a precision of $2.7\%$. These numbers are in excellent agreement with the results of Ref.~\cite{Forslund:2022xjq}: $0.77\%\,(2.6\%)$ at $2 E_{\rm b} = 10\,(3)\,$TeV. 

\renewcommand{\tabcolsep}{6pt}
\renewcommand{\arraystretch}{1.4}
\begin{table}[t]
\centering
\begin{tabular}{ lccc }    
\hfil  & cuts in Eq.~\eqref{eq:cuts_bb}& without $M_{bb}$ cut  \\\hline
$\mu^+\mu^- (h\rightarrow b\bar{b})$  & $2.0 \cdot 10^4$  & $4.2 \cdot 10^4$    \\
$\mu^+\mu^- b\bar{b}$, no $h$&  $4.5 \cdot 10^2$ & $5.8 \cdot 10^3$   \\
$\mu^+\mu^- t \bar{t}$ &  $3.6\cdot 10^2$ & $2.4 \cdot 10^3$    \\
$\mu^\pm \nu W^\mp h$ &  $1.1 \cdot 10^3$ & $2.8\cdot 10^3$  \\
$\mu^\pm \nu t b$ & $2.2\cdot 10^2$  & $1.6\cdot 10^3$  \\\hline
$(\Delta \sigma / \sigma_{\rm SM})^{ZZ\rightarrow h\rightarrow b\bar{b}}_{68\%}$ &  $0.75\%$ & $0.56\%$  \\
$S_{\rm SM}/B$ & $9.2$  & $3.3$ 
\end{tabular}\caption{Number of events expected for the SM $ZZ$ fusion Higgs production and decay to $b \bar{b}$, and for the dominant backgrounds, at the 10~TeV muon collider. In addition to the acceptance cuts~(\ref{eq:acceptance_cuts}), the selections in Eq.~\eqref{eq:cuts_bb} are applied in the left column, while the cut on $M_{bb}$ has been removed in the right column. The expected relative precision on the signal strength is defined as $(\Delta \sigma / \sigma_{\rm SM}) ^{ZZ\rightarrow h\rightarrow b\bar{b}}_{68\%} = \sqrt{B + S_{\rm SM}} / S_{\rm SM}$.}
\label{tab:hbb_results}
\end{table}

\begin{figure}[t]
\centering
\includegraphics[width=0.49\textwidth]{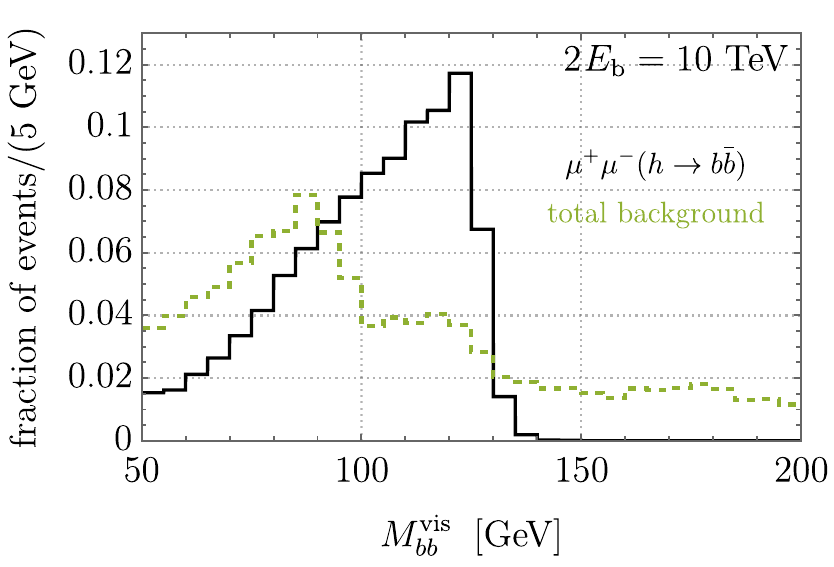}
\caption{\label{fig:mbb} Normalised distributions of the $b\bar{b}$ invariant mass for the Higgs production process and the total background, after applying the cuts in Eq.~\eqref{eq:acceptance_cuts} and requiring exactly 2 $b$-jets with $p_\perp^b > 30$~GeV. Here we denote the ${b\bar{b}}$ invariant mass as $M_{bb}^{\rm{vis}}$ to emphasise that it is constructed using only visible particles, without correcting for the invisible neutrinos produced in the decay of the bottom quarks.}
\end{figure}

However, by inspecting the $M_{bb}$ distribution in Fig.~\ref{fig:mbb} we notice that while selecting the $[100, 150]~\text{GeV}$ mass window like in Eq.~(\ref{eq:cuts_bb}) enhances the signal purity $S/B$, it also rejects a relatively large fraction of the signal, suggesting that applying this cut might not be optimal for the sensitivity to the Higgs signal strength. This is confirmed by the second column of Table~\ref{tab:hbb_results}. By eliminating the $M_{bb}$ cut we obtain a more precise determination of the signal strength, namely
\begin{equation}\label{eq:xsms}
    ( \Delta \sigma  / \sigma_{\rm SM} )^{ZZ\rightarrow h\rightarrow b\bar{b}}_{68\%} = \begin{cases}
        0.56\%\quad \, \,\,10\text{ TeV}\\
        2.3\% \quad\, \,\,\,\,\,\,\,\,3\text{ TeV}
    \end{cases}.
\end{equation}
The looser selection cuts produce a lower signal purity, $S/B = 3.3$, but this does not exclude that employing the extended region without the $M_{bb}$ cut could eventually be beneficial in order to access the $CP$-violating component of the Higgs production process, as it happens for the signal strength determination. 

We thus define the search region for the analysis of Section~\ref{sec:CPHTesting} without including the $M_{bb}$ cut, and discuss the impact on the results of performing the additional $M_{bb}$ selection.

It should be noted that our $M_{bb}$ variable---denoted as $M_{bb}^{\rm{vis}}$ in Fig.~\ref{fig:mbb}---is the invariant mass of the two $b$-tagged jets as reconstructed from the visible particles in the jets. The energy that is lost in the neutrinos from the $b$-quark decay is responsible for the tail on the left of the Higgs peak. It is possible to recover part of this energy by applying corrections to the $b$-jet momenta, thus reducing the tail. This could improve the signal acceptance of the $M_{bb}$ cut and change our conclusion on the relevance of this cut, provided that the corrections to the $b$-jet momenta do not push also the background to larger $M_{bb}$. Reference~\cite{Forslund:2022xjq} employed $b$-jet energy corrections and obtained the same result as ours using the $M_{bb}$ cut (which is weaker than the one we find without the cut). This suggests that $b$-jet energy corrections are not relevant for the present analysis, but a detailed full-simulation study based on a consolidated detector design would be needed in order to draw firm conclusions.

On top of the selections described above, in the study of the Higgs $CP$ properties of Section~\ref{sec:CPHTesting} we also impose an upper cut $p_\perp^\mu < 500$~GeV on the muon transverse momentum, for the following reasons. The kinematic configuration where one of the final state muons has large $p_\perp$ is sensitive to the $Z\mu \to h\mu$ scattering process at high momentum transfer, where one of the initial state muons undergoes a hard collision with an effective $Z$ boson emitted from the other muon. This process gives a small contribution in the SM, but is enhanced by the energy-growing effects of the BSM couplings in Eq.~\eqref{eq:hZZ}. Since we target a low-energy determination of the Higgs couplings in the VBF topology, we eliminate this potential BSM contribution via the $p_\perp^\mu < 500$~GeV cut.\footnote{An analysis targeting the $Z\mu \to h\mu$ hard scattering could be useful for characterising a potential discovery, but its BSM sensitivity is inferior to the low-energy probe.} Including this cut, the cross section for Higgs production in the SM is $\sigma_{h, \mathrm{SM}} = 3.9\;\mathrm{fb}$ (corresponding to $3.9\cdot 10^4$ events) while for the sum of the backgrounds we find $\sigma_{\rm bkg} = 1.0\;\mathrm{fb}$ ($1.0\cdot 10^4$ events).

\subsection{Testing the \boldmath{$CP$} property of the \boldmath{$hZZ$} coupling}\label{sec:CPHTesting}

We saw in Section~\ref{sec:CPVHiggsTheory} that a good sensitivity to $\tilde{c}_{zz}$ is offered by the variable 
\begin{equation}\label{eq:dphidef}
    \Delta \phi = \phi_{+} - \phi_{-}\,,
\end{equation}
where $\phi_+$ and $\phi_-$ are the $\mu^+$ and $\mu^-$ azimuthal angles, respectively. This variable is fully analogous to the ``signed $\Delta \phi_{jj}$'' that is employed at hadron colliders for the study of the $CP$ properties of the $hVV$ couplings in VBF production~\cite{Hankele:2006ma}. Notice that $\Delta \phi$, as defined in Eq.~\eqref{eq:dphidef}, ranges from $-2\pi$ to $+2\pi$ but of course its distribution is periodic with period $2\pi$, so that we can represent it in the range $[-\pi,+\pi]$.\footnote{Namely, we  define $\Delta \phi \equiv \phi_{+} - \phi_{-} - 2\pi\, \Theta (| \phi_{+} - \phi_{-}| - \pi)\, \mathrm{sgn}\,(\phi_{+} - \phi_{-})$, where $\Theta$ is the Heaviside step function. This differs from the variable $\Delta \phi_{\mu\mu}$ used in the other analyses of this work and in Ref.~\cite{Ruhdorfer:2023uea}: $\Delta \phi_{\mu\mu}= |\Delta\phi| = \mathrm{min}\, \big(|\phi_{+} - \phi_{-}|,\, 2\pi - |\phi_{+} - \phi_{-}|\big)$, ranging from $0$ to $\pi$. $\Delta \phi_{\mu\mu}$ is insensitive to $CP$ violation.}

\begin{figure}[t]
\centering
\includegraphics[width=0.48\textwidth]{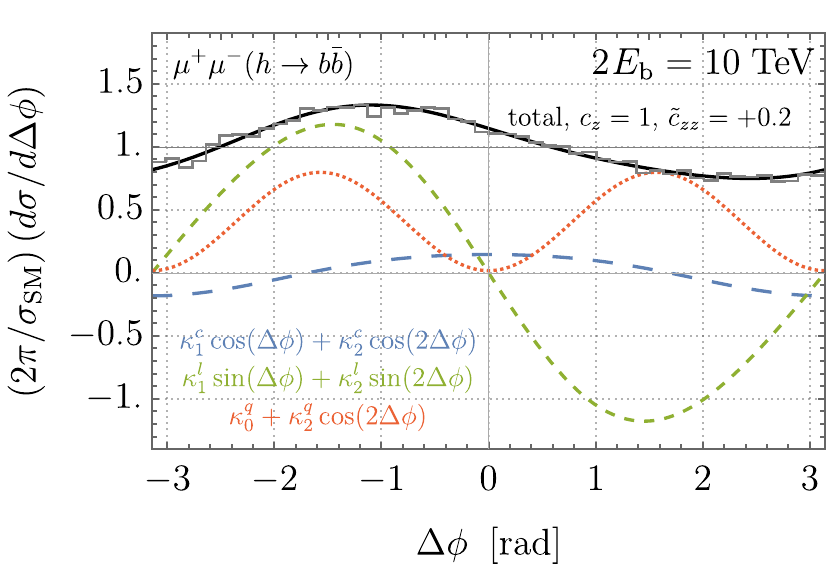}
\caption{\label{fig:CPVdelPhi} Normalised distributions of $\Delta\phi$ for the $\mu^+\mu^- (h\rightarrow b \bar{b})$ Higgs signal, after applying the event selection described in Section~\ref{sec:CPHTesting}. We show separately the three terms in square parentheses in Eq.~\eqref{eq:dspp}, obtained via a fit to reweighted MC events: the piece that multiplies $c_z^2$ after subtracting the trivial constant term (short-dashed red); the piece that multiplies $c_z \tilde{c}_{zz}$ (medium-dashed green); and the piece that multiplies $\tilde{c}_{zz}^2$ (long-dashed blue). The solid black line shows the total prediction for $c_z = 1$ and $\tilde{c}_{zz} = + 0.2$ (this large value of $\tilde{c}_{zz}$ was chosen for ease of illustration); in gray we show the histogram obtained from an independent MC sample, demonstrating good agreement.}
\end{figure}

The functional form of Eq.~\eqref{eq:dspp} is found to provide an excellent description of the $\Delta\phi$ distribution for the $\mu^+ \mu^- (h \to b\bar{b})$ production process, as shown in Fig.~\ref{fig:CPVdelPhi}. By means of event reweighting in \texttt{MadGraph}~\cite{Mattelaer:2016gcx} we isolate the three terms shown on separate lines of Eq.~\eqref{eq:dspp}. Then, we perform a fit to the detector-level $\Delta\phi$ distribution of each term and we extract the $\kappa_i^{c,l,q}$ coefficients reported in Table~\ref{tab:delPhi_fit}. The results are for the $2E_{\rm b} = 10$~TeV energy muon collider, and they include the event selection cuts described in Section~\ref{sec:CPVHiggsSignal}, without the $M_{bb}$ cut but including the $p_\perp^\mu < 500$~GeV requirement. The small dependence of $\mathrm{BR}(h\to b\bar{b})$ on $c_z$ and $\tilde{c}_{zz}$ is neglected in this procedure.

%


Some comments on the results of the fit are in order. First, in Fig.~\ref{fig:CPVdelPhi} we observe that the SM contribution (long-dashed blue line) is peaked at $\Delta\phi = 0$. This is due to the cut on $p_\perp^b$, which suppresses more the configurations where the Higgs is produced nearly at rest (which are preferentially those where the muons transverse momenta are back-to-back: $|\Delta \phi| \approx \pi$) compared to those where the Higgs has a sizeable momentum ($|\Delta\phi| \ll \pi$). Second, the piece that multiplies $c_z \tilde{c}_{zz}$ (medium-dashed green) is dominated by its $\sin(\Delta\phi)$ dependence, with only a small component of $\sin(2\Delta\phi)$. This quantitatively justifies our choice---see below---of the forward-backward asymmetry as $CP$ discriminant. Third, the term that multiplies $\tilde{c}_{zz}^2$ (short-dashed red) is proportional to $1 - \cos(2\Delta\phi)$ to a good approximation.

Since the contribution to the cross section that is linear in $\tilde{c}_{zz}$ is approximately proportional to $\sin(\Delta\phi)$, nearly-optimal sensitivity to $CP$ violation is obtained by the measurement of the forward-backward asymmetry 
\begin{equation}
{{A}}_{fb}^{\Delta \phi} \equiv \frac{\sigma_{f} - \sigma_{b}}{\sigma_{\rm{tot}}}\,,
\end{equation}
where $\sigma_{f(b)}$ is the cross section with positive (negative) $\Delta\phi$ in the region defined by the selection cuts, and ${\sigma_{\rm{tot}}}=\sigma_{f} + \sigma_{b}$ is the total cross section.

The expectation value of ${{A}}_{fb}^{\Delta \phi}$ is effectively zero in the SM, because $CP$ violation is small. The BSM prediction for ${{A}}_{fb}^{\Delta \phi}$ is readily obtained by integrating the differential cross section for Higgs production~\eqref{eq:dspp}, obtaining
\begin{equation}\label{eq:afb}
    A_{fb}^{\Delta \phi} = \frac{2}{\pi} \frac{
    {\sigma_{h,\rm{SM}}}}
    {\sigma_{\rm{tot}}}\,  \kappa_1^l c_z \tilde{c}_{zz}\;,
\end{equation}
where ${\sigma_{h,\rm{SM}}} = 3.9\;\mathrm{fb}$ is the SM Higgs production cross section while ${\sigma_{\rm{tot}}}$ is the total cross section including BSM effects and all the backgrounds listed in Table~\ref{tab:hbb_results}. Namely,
\begin{equation}\label{eq:stot}
{\sigma_{\rm{tot}}}
= (c_z^2+\kappa_0^q \tilde{c}_{zz}^2){\sigma_{h,\rm{SM}}}
+{\sigma_{\rm{bkg}}}
\;,
\end{equation}
where $\sigma_{\rm{bkg}} = 1.0\;\mathrm{fb}$. It should be noted that the background process $\mu^\pm \nu W^\mp h$ is sensitive to the $CP$-odd BSM coupling $\tilde{c}_{zz}$ and it also generates a contribution to the asymmetry, which is however found to be negligible.\footnote{Under our assumptions of vanishing BSM corrections to the $h\gamma\gamma$ and $hZ\gamma$ couplings, namely $\tilde{c}_{\gamma\gamma} = \tilde{c}_{z\gamma} = 0$, the $\mu^\pm \nu W^\mp h$ production process is sensitive to $\tilde{c}_{zz}$ through both the $hWW$ and $hZZ$ vertices, because Eq.~\eqref{eq:ctildeww} dictates $\tilde{c}_{ww} = \tilde{c}_{zz}$.} 


\renewcommand{\tabcolsep}{6pt}
\renewcommand{\arraystretch}{1.4}
\begin{table}[t]
\centering
\begin{tabular}{ cccccc }    
$\kappa_1^c$  & $\kappa_2^c$ & $\kappa_1^l$ & $\kappa_2^l$  &   $\kappa_0^q$    &   $\kappa_2^q$  \\\hline
$0.16$  & $-\,0.019$ & $-\,1.17 $ & $-\,0.071$  &   $0.41$    &   $-\,0.39$ 
\end{tabular}\caption{Coefficients of the differential cross section in $\Delta\phi$ for the $\mu^+ \mu^-\to \mu^+\mu^- (h\to b\bar{b})$ process, Eq.~(\ref{eq:dspp}). They are obtained through a fit to the detector-level $\Delta\phi$ distribution after the selection described in Section~\ref{sec:CPVHiggsSignal}, without the $M_{bb}$ cut but including the $p_\perp^\mu < 500\;\mathrm{GeV}$ requirement. The MC uncertainties on the values of $\kappa_1^l$ and $\kappa_0^q$, which are used to set our constraints, are around $1\%$ and safely negligible.}
\label{tab:delPhi_fit}
\end{table}

We estimate the error on the ${{A}}_{fb}^{\Delta \phi}$ measurement as the sum in quadrature of a statistical and a systematic component
\begin{equation}
\Delta {{A}}_{fb}^{\Delta \phi}
=\sqrt{
\frac1{N_{\rm tot}^{\textsc{sm}}}+\varepsilon^2}\,,
\end{equation}
where $N_{\rm tot}^{\textsc{sm}}= 4.9\, \cdot \, 10^4$ is the total number of expected events with $10~{\textrm{ab}}^{-1}$ integrated luminosity. It is difficult to guess the physical origin of the dominant systematic uncertainties and to estimate the error $\varepsilon$. Common sources of uncertainties such as the ones on the luminosity, reconstruction efficiencies and acceptance, cancel out in the measurement of the asymmetry. The uncertainty should emerge from some mismodeling of the detector response that affects the $\Delta\phi>0$ configurations, where the $\mu^+$ has a larger azimuthal angle than the $\mu^-$, differently from the $\Delta\phi<0$ configuration with switched $\mu^+$ and $\mu^-$ transverse momenta. Since it is hard to identify such asymmetric effects, the systematic uncertainty is arguably small. Nevertheless, in what follows we illustrate the dependence on $\varepsilon$ by considering the three values $0,\, 0.5$ and $1\%$.

We quantify the sensitivity in terms of the $95\%$~CL exclusion limit under the hypothesis of the observation of a vanishing (SM-like) asymmetry. The allowed region in the parameter space is defined by the relation
\begin{equation}
\big|{{A}}_{fb}^{\Delta \phi}\big|<
{1.96}\;\Delta {{A}}_{fb}^{\Delta \phi}\,.
\end{equation}
The limit is set in a region of the parameter space that is close to the SM point $({{c}}_{z}=1,{\tilde{c}}_{zz}=0)$, where the BSM contribution to the total cross section in Eq.~(\ref{eq:stot}) is negligible. The asymmetry in Eq.~(\ref{eq:afb}) is approximately $A_{fb}^{\Delta \phi} \approx -\, 0.60 \, c_z \tilde{c}_{zz}$, and the constraints are 
\begin{equation}\label{eq:bounds_ctzz}
     | c_z \tilde{c}_{zz} | <  \{1.5,\, 2.2 ,\, 3.6 \} \cdot 10^{-2}\,,\;\; \varepsilon= \{0,\, 0.5,\, 1\}\%\,.
\end{equation}
These results are obtained with the selections described in Section~\ref{sec:CPVHiggsSignal}, excluding the $M_{bb}$ cut which, on the other hand, would improve the signal purity. We verified that the $M_{bb}$ cut does not improve the sensitivity if $\varepsilon \lesssim 1\%$, but it would instead be beneficial if the systematic uncertainty was larger.

For $c_z=1$, the projected constraints in Eq.~(\ref{eq:bounds_ctzz}) on the $CP$-odd coefficient $\tilde{c}_{zz}$ are almost two orders of magnitude stronger than the current bounds from ATLAS and CMS. In order to compare with the sensitivity of future collider projects, we use Eq.~\eqref{eq:cZZvsfCP} to translate our results into bounds on the $f_{CP}^{hZZ}$ parameter that is employed in Ref.~\cite{Gritsan:2022php}, obtaining
\begin{equation}
     f_{CP}^{hZZ} < \{0.64,\, 1.4,\, 3.8\}\cdot 10^{-6}\,,\;\; \varepsilon= \{0,\, 0.5,\, 1\}\%\,. 
\end{equation}
At the HL-LHC, the expected $95\%$~CL sensitivity is $f_{CP}^{hZZ} < 10\cdot 10^{-6}$, mainly driven by electroweak Higgs production followed by $h\to \tau^+ \tau^-$~\cite{CMS:2022uox,CMS:2022uox_suppl}. At an $e^+ e^-$ collider with $2\hspace{0.2mm}E_{\rm b} = 250, 350, 1000$~GeV and $L=250, 350, 1000$~fb$^{-1}$, respectively, one expects the $95\%$ CL constraints $f_{CP}^{hZZ} < (160, 120 , 12)\cdot 10^{-6}$ via the $e^+ e^- \to Zh$ process~\cite{Gritsan:2022php} (see also Ref.~\cite{Vukasinovic:2024ftu} for initial results on $ZZ$ fusion at 1 TeV). A recent study shows that beam polarisation can improve the $e^+ e^- \to Zh$ bound to $f_{CP}^{hZZ} < 18\cdot 10^{-6}$ at 250 GeV~\cite{Li:2024mxw}. We conclude that a $10\;$TeV muon collider is one order of magnitude more sensitive to $f_{CP}^{hZZ}$ than other future colliders for a systematic uncertainty of less than $0.5\%$ in the measurement of the asymmetry.

It is useful to evaluate the impact of our projected bounds~\eqref{eq:bounds_ctzz} on concrete UV-complete models. Models containing new vector-like fermions are a relevant class of UV completions where $CP$-violating $hVV$ interactions can be generated. As an illustration, we consider a model with two vector-like lepton multiplets, $\chi_2 \sim \mathbf{2}_{1/2}$ and $\chi_3 \sim \mathbf{3}_{1}$ under SU(2)$_L\times$U(1)$_Y$, with common mass $M_\chi$ and coupled to the SM Higgs doublet via $CP$-violating Yukawas $y_{\chi_L}$ and $y_{\chi_R}$. Integrating out the new fermions at one loop~\cite{Bakshi:2021ofj} results in
\begin{equation}
\tilde{c}_{zz} = -\, \frac{\mathrm{Im} (y_{\chi_L} y_{\chi_R}^\ast) v^2}{4\pi^2 M_\chi^2} \Big[ c_w^4 + \frac{5}{3} s_w^2 c_w^2 + 4 s_w^4 \Big]\,,
\end{equation}
where $s_w$ and $c_w$ are the sine and cosine of the weak mixing angle. Our most optimistic bound in Eq.~\eqref{eq:bounds_ctzz} translates to $M_\chi \gtrsim 340\;\mathrm{GeV}\, |\mathrm{Im} (y_{\chi_L} y_{\chi_R}^\ast)|^{1/2}$, at $95\%$~CL. Thus, provided the BSM Yukawa couplings of this fermionic completion have absolute values $|y_{\chi_{L,R}}| \gtrsim 1$ and an $\mathcal{O}(1)$ physical phase, the muon collider will be able to meaningfully test $CP$ violation in the low-energy EFTs arising from this class of UV completions.

\begin{figure}[t]
\centering
\includegraphics[width=0.4825\textwidth]{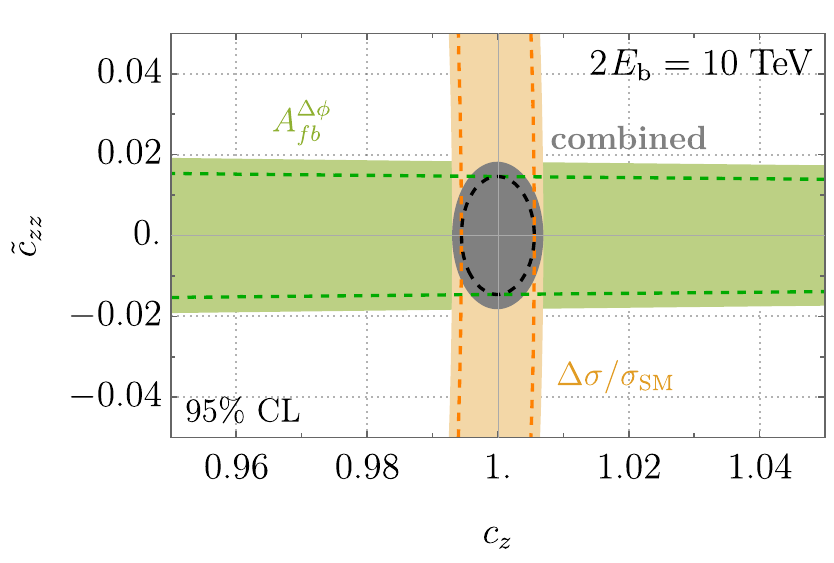}
\caption{\label{fig:cz_ctzz} Projected constraints on the coefficients parametrising $CP$-even ($c_z$) and $CP$-odd ($\tilde{c}_{zz}$) BSM effects in the $hZZ$ interaction. We show in orange the measurement of only the inclusive cross section; in green, the forward-backward asymmetry alone; in gray, the combination of both observables. Only statistical uncertainties are included. Shaded regions (dashed contours) are drawn using $\chi^2$ levels of $\Delta\chi^2=5.99$, corresponding to two parameters~($\Delta\chi^2=3.84$, one parameter).}
\end{figure}

Finally, we combine the measurement of the forward-backward asymmetry $A^{\Delta\phi}_{fb}$ with that of the Higgs production cross section in order to derive exclusion contours in the $(c_z, \tilde{c}_{zz})$ plane. The cross section is 
\begin{equation}
(\Delta \sigma / \sigma_{\rm SM})^{ZZ\rightarrow h\rightarrow b\bar{b}}= c_z^2 + \kappa_0^q \tilde{c}_{zz}^2 - 1\;,
\end{equation}
and the error on the measurement, estimated as in Section~\ref{sec:CPVHiggsSignal}, is\footnote{The result is not identical to the one in Eq.~(\ref{eq:xsms}) because it now includes the $p_\perp^\mu < 500$~GeV cut.} 
\begin{equation}
(\Delta \sigma / \sigma_{\rm SM})^{ZZ\rightarrow h\rightarrow b\bar{b}}_{68\%} = 0.57\%\,.
\end{equation}
By combining this with the measurement of the asymmetry via a simple $\chi^2$, we obtain the contours reported in Fig.~\ref{fig:cz_ctzz}. 

The result of the combined fit outlines the optimal complementarity of the two measurements for the simultaneous determination of the $CP$-even and $CP$-odd components of the $hZZ$ coupling. The combined measurement would also allow us to distinguish a non-SM $CP$-even coupling from a $CP$-odd component, potentially enabling the discovery of $CP$ violation in the $hZZ$ interaction. We also remark that, as expected, the measurement of the asymmetry dominates the single-parameter constraint on the $CP$-odd coefficient ${\tilde{c}}_{zz}$: if only the cross section was measured, the $95\%$~CL bound would degrade by one order of magnitude ($|\tilde{c}_{zz}| < 0.17$), becoming less strong than the HL-LHC projections.

\section{Conclusions}\label{sec:conc}

We investigated the opportunities offered by the detection of  forward muons at a muon collider. Our results provide a strong physics case for the design of a forward muon detector, as well as a set of benchmarks for the assessment of its required performances. We also outlined the interplay with the main detector configuration for the selection of invisible final states, in particular with the capability of the main detector to veto visible objects with the widest possible energy and angular coverage.

In Section~\ref{sec:inclusive}, we confirmed and in some cases slightly strengthened the sensitivity projections for the measurement of the inclusive Higgs production cross section in $ZZ$ fusion and of the Higgs to invisible decay branching ratio. The results depend on assumptions on the performances of the forward muon detector, such as the resolution on the muon energy. With our benchmark resolution of $10\%$, a $5$ per mille measurement of the inclusive Higgs cross section is possible. This translates to an absolute measurement of the $hZZ$ coupling that is nearly as precise as the best determination that can be envisaged at other future colliders. The Higgs to invisible branching ratio can be probed at the $1$ per mille level with benchmark detector parameters, which is comparable to future $e^+e^-$ colliders. With our most optimistic muon energy resolution of $1\%$, the expected sensitivity is a factor of 2 weaker than the current FCC-hh projection.

In Section~\ref{sec:BSM}, we studied new invisible particles with Higgs portal couplings. In comparison with a previous study, we find less strong exclusions after considering all the relevant backgrounds and a more realistic simulation of the BES and the forward detector response. Even with this degradation, the muon collider sensitivity compares favourably---see Fig.~\ref{fig:portalComp}---to other direct or indirect future probes of this BSM physics scenario. The sensitivity is strongly affected by the assumed muon energy resolution, but also by the angular coverage of the main detector that is used to veto visible objects. It should be noted that our optimistic main detector coverage $\theta_{\rm{MD}}=5^\circ$ is in fact less ambitious than the aspirational IMCC target~\cite{InternationalMuonCollider:2024jyv} $\eta_{\rm{MD}}=4$, i.e. $\theta_{\rm{MD}}=2.1^\circ$.

In Section~\ref{sec:hZZ_CP}, we illustrated the advantages of observing the forward muons and measuring their azimuthal angles in order to access the quantum-mechanical interference between the exchange of vector bosons with different helicities in $ZZ$-scattering or -fusion processes. On general grounds, this provides the muon collider with novel unexplored capabilities to resurrect the interference effects among initial-state vector bosons, and not only to study final-state interference by measuring angular correlations of the decay products. While this observation could open the door to a rich program of studies in several VBS or VBF processes, for a simple illustration we considered the determination of the $CP$ properties of the $hZZ$ coupling in $ZZ$-fusion Higgs production. We found an excellent potential sensitivity to a BSM $CP$-odd component of the coupling, surpassing the expectations for future $e^+ e^-$ colliders. In contrast to the measurement of the inclusive Higgs production cross section and the analyses of final states containing invisible particles, interference resurrection only requires information on the (azimuthal) angles of the forward muons. It is therefore largely insensitive to the resolution that can be achieved in the measurement of the forward muon energies.

We can summarise our results also from the viewpoint of the forward detector capabilities that are needed to accomplish them. A basic detector would just tag the presence of the muons in the forward region. This would make it possible to distinguish neutral from charged vector boson processes, enabling in particular the independent determination of the Higgs couplings to the $W$ and to the $Z$, as well as improving the characterisation of VBS processes. On the other hand, a basic muon tagger would not enable any of the studies presented in this paper. In particular, while the tagger would offer one handle to identify the production of invisible BSM particles, we saw in Section~\ref{sec:BSM} that the measurement of the muon energy is also required for the suppression of the otherwise overwhelming background.

A more advanced detector would measure the trajectory of the muons, but not their energy. This would enable the precise determination of the $CP$ properties of the $hZZ$ coupling, as in Section~\ref{sec:hZZ_CP}. It should be noted that even a modest accuracy on the measurement of the azimuthal angle would be sufficient for this task: the sensitivity comes from a modulation of the $\Delta\phi$ distribution---see Fig.~\ref{fig:CPVdelPhi}---that varies slowly with $\Delta\phi$. For a resolution in the measurement of the angle that is better than around $1$ radian, the corresponding smearing would not deform the distribution significantly and would not affect the sensitivity dramatically.

A complete forward detector would also measure the energy of the muons, enabling all the studies presented in this paper. With our benchmark energy resolution of $10\%$, two important targets can be achieved. One is the absolute determination of the Higgs couplings to a level that is comparable with future low-energy Higgs factories. The other is a sensitivity to the Higgs to invisible branching ratio that approaches the SM value, potentially enabling the observation of the Higgs decaying to invisible particles even within the SM. Since our results depend quite strongly on the resolution, these targets would be missed if the resolution was worse than $10\%$. Conversely, there is a strong physics case for lowering the resolution below $10\%$. This would improve not only the precision of Higgs measurements, but also the sensitivity to new particles produced through the Higgs portal, especially if the reduction of the energy uncertainties is accompanied by an extension of the angular coverage of the main detector.

\begin{acknowledgments}
This work has been endorsed by the IMCC. We acknowledge valuable feedback on the draft by the IMCC Publication Speakers Committee. We thank Samuel Homiller and Simon Knapen for useful discussions on muon collider physics. We are grateful to Uli Haisch for sharing with us numerical data from Ref.~\cite{Haisch:2022rkm}. MR was supported in part by the NSF grant PHY-2014071 and by a Feodor--Lynen Research Fellowship awarded by the Humboldt Foundation. MR was also supported by NSF Grant PHY-2310429, Simons Investigator Award No.~824870, DOE HEP QuantISED award \#100495, the Gordon and Betty Moore Foundation Grant GBMF7946, and the U.S.~Department of Energy (DOE), Office of Science, National Quantum Information Science Research Centers, Superconducting Quantum Materials and Systems Center (SQMS) under contract No.~DEAC02-07CH11359. ES was supported by the Science and Technology Facilities Council under the Ernest Rutherford Fellowship ST/X003612/1. AW was supported by the grant PID2023-146686NB-C31 funded by MCIU/AEI/10.13039/501100011033 and by FEDER, UE. The work of MR and ES was performed in part at the Aspen Center for Physics, which is supported by National Science Foundation grant PHY-2210452 and by a grant from the Simons Foundation (1161654, Troyer).
\end{acknowledgments}

\appendix

\section{\boldmath{$CP$} violation in \boldmath{$hVV$} couplings}\label{sec:CP_details}

The $CP$-violating interactions of the Higgs with vector bosons are parametrised in the Higgs basis for the dimension-6 SMEFT as~\cite{Azatov:2022kbs}
\begin{align}\label{eq:ctHB}
&\,\mathcal{L}_{hVV}^{\slashed{\textsc{cp}}} = \frac{h}{v}\bigg[ \tilde{c}_{zz} \frac{g^2+g^{\prime\, 2}}{4} Z_{\mu\nu} \widetilde{Z}^{\mu\nu} + \tilde{c}_{z\gamma} \frac{g g^\prime}{2}Z_{\mu\nu} \widetilde{A}^{\mu\nu} \nonumber \\ 
+&\, \tilde{c}_{\gamma\gamma} \frac{g^2 g^{\prime\,2}}{4 (g^2 + g^{\prime\,2})} A_{\mu\nu} \widetilde{A}^{\mu\nu} + \tilde{c}_{ww} \frac{g^2}{2} W_{\mu\nu}^+ \widetilde{W}^{\mu\nu -} \bigg]\,,
\end{align}
where $g$ and $g^\prime$ are the SU$(2)_L$ and U$(1)_Y$ gauge couplings. The coefficient $\tilde{c}_{ww}$ is not independent from the other three,
\begin{equation}\label{eq:ctildeww}
\tilde{c}_{ww} = \tilde{c}_{zz} + \frac{2 g^{\prime\,2}}{g^2 + g^{\prime\,2}} \,\tilde{c}_{z\gamma} + \frac{g^{\prime\,4}}{(g^2 + g^{\prime\,2})^2} \,\tilde{c}_{\gamma\gamma}\,.
\end{equation}
In terms of Wilson coefficients in the Warsaw basis~\cite{Grzadkowski:2010es},
\begin{align}
    \mathcal{L}_{\rm Warsaw}^{\slashed{\textsc{cp}}} \supset \frac{\tilde{c}_{WW}}{v^2}|H|^2 W^a_{\mu\nu}&\widetilde{W}^{a\, \mu\nu} + \frac{\tilde{c}_{WB}}{v^2}H^\dagger \sigma^a H \widetilde{W}^a_{\mu\nu} B^{\mu\nu} \nonumber \\
    &+\frac{\tilde{c}_{BB}}{v^2}|H|^2 B_{\mu\nu} \widetilde{B}^{\mu\nu}\,,
\end{align}
we have the identifications
\begin{align}
    \tilde{c}_{zz} =&\; 4\,\frac{g^2\, \tilde{c}_{WW} + g\, g'\, \tilde{c}_{WB} + g'^2\, \tilde{c}_{BB}}{(g^2 + g'^2)^2}\,, \nonumber \\
        \tilde{c}_{z\gamma} =&\; 2\,\frac{2\, \tilde{c}_{WW} - \frac{g^2 - g^{\prime\,2}}{g g^\prime} \tilde{c}_{WB} -2\, \tilde{c}_{BB}}{g^2 + g'^2}\,, \\
        \tilde{c}_{\gamma\gamma} =&\; 4\, \Big( \frac{\tilde{c}_{WW}}{g^2} - \frac{\tilde{c}_{WB}}{g g^\prime} + \frac{\tilde{c}_{BB}}{g^{\prime\,2}} \Big) \,. \nonumber
\end{align}
In this work we assume $\tilde{c}_{z\gamma}=\tilde{c}_{\gamma\gamma}=0$, which corresponds to the occurrence of a cancellation among the three Warsaw basis $CP$-odd Wilson coefficients. As discussed in the main text, it would be interesting to generalise the analysis and assess the muon collider potential to distinguish the three coefficients by a combined study of $ZZ$-, $Z\gamma$- and $\gamma\gamma$-fusion amplitudes and their interference.

Another commonly-used parametrisation employs the coefficients of the three-point amplitudes. For the $hZZ$ amplitude
\begin{align}
     v\hspace{0.1mm} \mathcal{A}(h \to Z_1 Z_2) = a_1^{h ZZ} & m_Z^2 \, \epsilon_1^{* \mu} \epsilon_{2\mu}^* + a_2^{h ZZ} f_{\mu\nu}^{*\, (1)} f^{*\, (2) \mu\nu} \nonumber\\
    &\quad+ a_3^{h ZZ} f_{\mu\nu}^{*\, (1)}\tilde{f}^{*\, (2) \mu\nu}\,,
\end{align}
where $f^{(i)}_{\mu\nu} = p_{i\, \mu} \epsilon_{i\, \nu} - p_{i\, \nu} \epsilon_{i\, \mu}$ and $\tilde{f}^{(i)}_{\mu\nu} = \epsilon_{\mu\nu\rho\sigma} f^{(i)\rho\sigma}/2$.

The amplitude coefficient $a_2^{h ZZ }$ corresponds to a $CP$-even interaction of the Higgs with the $Z$ boson field strengths, which we set to zero. The $CP$-violating component $a_3^{h ZZ }$ is in one-to-one correspondence with $\tilde{c}_{zz}$, namely 
\begin{equation}
a_3^{h Z Z} = -\,\frac{g^2+ g'^2}{2}\, \tilde{c}_{zz}\,.
\end{equation}
The other $CP$-even term is
\begin{equation}
a_1^{h Z Z} = 2\,c_{z}\,,
\end{equation}
with $c_{z}$ as in Eq.~(\ref{eq:hZZ}). In the SM, $c_{z}=1$.

It is also customary to quantify the amount of $CP$ violation in terms of partial Higgs decay widths, defining
\begin{equation}\label{eq:fCPZZ}
    f_{CP}^{hZZ} \equiv \frac{\Gamma_{h\rightarrow ZZ}^{CP\; \mathrm{odd}}}{\Gamma_{h\rightarrow ZZ}^{CP\;\mathrm{even}} + \Gamma_{h\rightarrow ZZ}^{CP\;\mathrm{odd}}}\;.
\end{equation}
In terms of the amplitude coefficients
\begin{equation}
    f_{CP}^{hZZ} = \frac{|a_3^{hZZ}|^2}{\tfrac{\Gamma_1}{\Gamma_3} |a_1^{hZZ}|^2 + |a_3^{hZZ}|^2}\;,
\end{equation}
where $\Gamma_{1(3)}$ is the decay width for $h\to ZZ^\ast$ with only $a_{1(3)}^{hZZ}=1$ turned on. The numerical value is (see e.g. Ref.~\cite{CMS:2021nnc}) $\Gamma_3/\Gamma_1 \approx 0.153$. 

We can thus express $f_{CP}^{hZZ}$ in terms of $\tilde{c}_{zz}$ and $c_{z}$ as
\begin{equation}\label{eq:cZZvsfCP}
    f_{CP}^{hZZ} \approx \frac{2.88\cdot 10^{-3}\,\tilde{c}_{zz}^2}{c_{z}^2 + 2.88\cdot 10^{-3}\,\tilde{c}_{zz}^2}\,.
\end{equation}

\bibliography{bibliography}

\end{document}